\documentclass[%
 aip,
 amsmath,amssymb,
 reprint,%
]{revtex4-1}

\usepackage{graphicx}% Include figure files
\usepackage{dcolumn}% Align table columns on decimal point
\usepackage{bm}% bold math
\usepackage[utf8]{inputenc}
\usepackage[T1]{fontenc}
\usepackage{mathptmx}
\usepackage{etoolbox}
\usepackage{colortbl}

\usepackage{graphicx}% Include figure files
\usepackage{dcolumn}% Align table columns on decimal point
\usepackage{bm}% bold math
\usepackage{caption}
\usepackage{subfigure}
\usepackage{subcaption}
\usepackage{upgreek}
\usepackage{subfloat}
\usepackage{float}
\usepackage{comment}
\usepackage{gensymb}

\usepackage{siunitx}
\usepackage{braket}
\usepackage{colortbl}
\usepackage{float}

\newcommand{\com}[1]{{\color{red}}}

\usepackage{booktabs} % For better-looking tables
\usepackage{tabularx} % Allows for dynamic column width
\usepackage{lipsum} 
\makeatletter
\def\@email#1#2{%
 \endgroup
 \patchcmd{\titleblock@produce}
  {\frontmatter@RRAPformat}
  {\frontmatter@RRAPformat{\produce@RRAP{*#1\href{mailto:#2}{#2}}}\frontmatter@RRAPformat}
  {}{}
}%
\makeatother
\begin{document}

\preprint{AIP/123-QED}

\title[]{High-quality entangled photon source \\ by symmetric beam displacement design}
%\begin{comment}
    \author{G. Paganini}
 \email{giacomo.paganini@icfo.eu}
 \affiliation{ICFO-Institut de Ciències Fotòniques, Mediterranean Technology Park, E-08860 Castelldefels (Barcelona), Spain}
 
\author{A. Cuevas}
\affiliation{ICFO-Institut de Ciències Fotòniques, Mediterranean Technology Park, E-08860 Castelldefels (Barcelona), Spain}

\author{R. Camphausen}
\affiliation{ICFO-Institut de Ciències Fotòniques, Mediterranean Technology Park, E-08860 Castelldefels (Barcelona), Spain}

\author{A. Demuth}
\affiliation{ICFO-Institut de Ciències Fotòniques, Mediterranean Technology Park, E-08860 Castelldefels (Barcelona), Spain}

\author{V. Pruneri}
\affiliation{ICFO-Institut de Ciències Fotòniques, Mediterranean Technology Park, E-08860 Castelldefels (Barcelona), Spain}
\affiliation{ICREA-Institució Catalana de Recerca i Estudis Avançats, E-08010 Barcelona, Spain
}
%\end{comment}

% word count including abstract, aknowledgments and authors = 3667 
% word count for APL = 3374

\date{\today}% It is always \today, today,
             %  but any date may be explicitly specified

\begin{abstract}
Entangled photon sources (EPSs) are pivotal in advancing quantum communication, computing and sensing. The demand for deploying efficient, robust EPSs in the field, characterized by exceptional (phase) stability, has become increasingly apparent. This work introduces a polarization-entangled photon source, leveraging type-0 spontaneous parametric down-conversion, and constructed using commercial bulk optomechanical components. Our system is versatile, enabling the generation of N00N states for sensing applications or Bell states for quantum key distribution protocols. We attained a maximal Bell inequality violation, with the average entanglement visibility exceeding 99\% . The potential for further performance enhancements is also explored.
\end{abstract}

\maketitle

\section{\label{sec:Intro}Introduction}

Photonic entanglement is a crucial resource for modern quantum science and technology, including quantum key distribution \cite{GisinRev,kimbleQuantumInternet2008}, teleportation \cite{telep,telep_zeil}, quantum-enhanced metrology \cite{sensing_rev, imaging_rev}, and quantum computing \cite{QuComp1, QuComp2}. In all these fields, the efficient generation of high quality entangled photon pairs is essential. For instance, the advancement of next-generation projects aimed at establishing a global quantum network, through ground-to-satellite and inter-satellite connections, necessitates the development of compact, robust, and durable sources of entangled photons with high brightness and entanglement fidelity \cite{micius_rev}.

Entangled states of light can be encoded in many degrees of freedom, such as frequency-bin \cite{freq_ent}, energy-time \cite{energy_time_ent}, time-bin \cite{time_bin_ent}, orbital angular momentum \cite{oam_entang}, and even in multiple degrees of freedom simultaneously, known as hyper-entanglement \cite{hyper_ent}. Polarization entanglement, however, has become one of the leading resources for quantum communication applications due to its relative ease of control, scalability, and compatibility with telecom infrastructure \cite{pol_ent}. Additionally, polarization-encoded quantum states have demonstrated high robustness when propagating through a turbulent atmosphere, making them the preferred choice for free-space and satellite quantum links \cite{micius_rev,entang_aspelmeyer,yinEntanglementbasedSecureQuantum2020a}.\\
Various technically well-developed physical platforms enable building entangled photon sources (EPS), both in bulk \cite{SPDC_EPS_rev}, and integrated photonics \cite{waveguide_entang,qdots_3,silicon_PIC}.
Integrated EPSs offer compactness, single mode characteristics, and especially high brightness, on the order of $\sim10^7 \, \si{pairs/s/mW}$ pump power in the case of PPLN waveguides \cite{Kuo:20}.
However, they currently still face technical challenges that hinder their use in realistic field-deployment, particularly, high out-coupling \cite{mahmudluFullyOnchipPhotonic2023} losses from EPS to fibre or free-space link.
In contrast, EPSs based on spontaneous parametric down-conversion (SPDC) in bulk optics can provide near-ideal heralding efficiencies \cite{SPDC_EPS_rev}, and broad spectral, and spatial multimode characteristics. They also allow access to various degrees of freedom, like position and momentum, which in turn enables manipulation of structured light like OAM and vector vortex modes. These features enable hyper-entanglement capabilities \cite{achatz_2023} and multi-party quantum communications \cite{WDM_2018, ortega_2021} while also facilitating higher heralding performance.
Therefore, combined with the wide availability and modularity of off-the-shelf components, at present bulk optics remain the platform of choice for EPS development.

A number of bulk optical geometries have been demonstrated to produce polarization entanglement via SPDC, including the crossed-crystal, Sagnac interferometer, and double-pass ("folded sandwich") schemes \cite{SPDC_EPS_rev}.\\
In recent years, EPS designs similar to linear double-path interferometers have emerged.
In these designs, a beam displacer (BD) splits the pump into two orthogonally and linearly polarized beams, deviating only one from the input trajectory while maintaining its original k-vector. By passing the pump beam parallel through either one \cite{linearDisp_EPS} or two \cite{dual_cryst_eps} non-linear crystals, parallel SPDC emission can be achieved, and then recombined by a second BD. This solution is easily scalable as all beams propagate in only one direction, and experiences less walk-off compared to the crossed-crystal scheme.

In this paper, we introduce a compact and robust scheme for polarization entanglement generation, which improves the existing beam displacer EPS designs through the novel combination of half-wave plates and Savart plates, the latter being a composition of birefringent plates that splits a beam in two parallel beams with orthogonal polarization. This innovative scheme intrinsically improves on longitudinal walkoff and compactness compared to previous work, without compromising on the mechanical sensitivity and on the quality of the generated entanglement.
Our device generates entangled photon pairs in the form of N00N states or $\ket{\Phi^{\pm}}$ Bell states, achieving near-ideal fidelity ($F=0.992\pm0.001$) and high brightness $CC = (9.50\pm0.03) \times 10^4 {\text{ pairs}}/{(\text{s}\cdot \text{mW})}$. It features a modular, miniaturizable architecture, making it suitable for both free-space and fiber-based quantum communication applications.

\section{\label{sec:Methods}Methods}
\normalcolor

\subsection{Conceptual Design}
\label{sec:design}

\begin{figure}[ht]
\begin{center}
\includegraphics[width=0.45\textwidth]{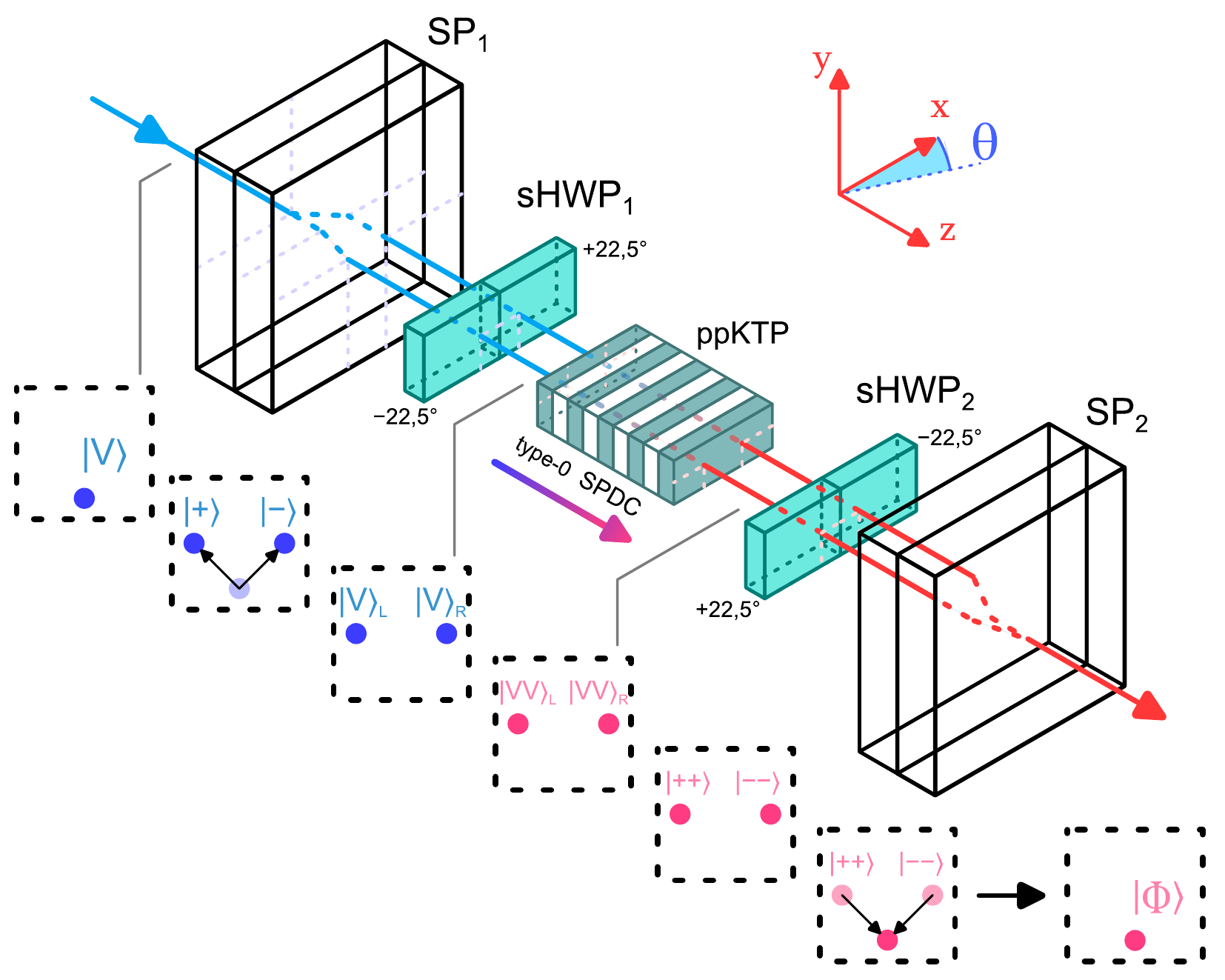}
\caption{\label{fig:EPS_design} \textbf{SPaDEs working principle.} $V$-polarized pump is split into $\ket{+}$ and $\ket{-}$ components by the first Savart Plate SP$_{1}$.
The segmented half-wave plate sHWP$_{1}$ rotates the polarization of both parallel modes to $\ket{V}$ (subscript L(R) indicates left(right) beam).
$\ket{VV}$ photon pairs are generated in ppKTP crystal, L(R) photons are rotated to $\ket{++}$($\ket{--}$) by sHWP$_{2}$, and recombined by SP$_{2}$.
Polarization(s) and transverse ($xy$) position(s) of state after each component shown in boxes below.
SP$_2$ is tilted along the $xz$-plane, i.e. $\theta$, as defined in coordinate axes (top right).}
\end{center}
\end{figure}

In this work, we present a high-performance bulk-optics design for a polarization EPS, named SPaDEs (Symmetric Parallel Displacement Entanglement Source). Its configuration, depicted in Fig.~\ref{fig:EPS_design}, draws inspiration from lateral displacement EPSs \cite{dual_cryst_eps, linearDisp_EPS, linear_eps}.

A key advancement is that the longitudinal walkoff between the two parallel spatial SPDC emission modes is eliminated at every cross-section along the propagation direction, not just at the output wavefront. This is achieved by substituting the birefringent single-beam displacement (BD) elements (used for both pump and SPDC photons) with the joint dual-beam displacement elements, referred to as balanced beam displacers (BBD). These components can be manufactured in both integrated and fiber optics versions, but this manuscript focuses on the bulk optics case. One example is the Savart plate (SP), consisting of two single-plate BDs whose axes are rotated $90\degree$ in the transverse plane relative to each other, with no gap between the two plates.
When aligned with displacement axes at $\pm45\degree$, an SP splits a vertically ($V$) polarized single-beam state with transverse spatial coordinates $(x,y)$, $\ket{V}\ket{x,y}$,
into a two-beams state $\frac{1}{\sqrt{2}}\left(\ket{+}\ket{x+\delta,y+\delta}+e^{i\theta}\ket{-}\ket{x-\delta,y+\delta}\right)$. That is, the two parallel output beams are separated along $x$ by a shear $2\delta$.
Here $\ket{\pm}\equiv \frac{1}{\sqrt{2}}(\ket{H}\pm\ket{V})$ and $\ket{H}$ represents the horizontal polarization.
Therefore, an SP functions similarly to a polarizing beam splitter (PBS) that maintains the k-vectors from input to output \cite{robin}.

Referring to the scheme in Fig.~\ref{fig:EPS_design}, the pump beam (blue) has vertical polarization. The first component, $\mathrm{SP_1}$, splits the beam into its $\ket{+}$ and $\ket{-}$ components. After that, a pair of half waveplates (HWPs) rotate each component into the vertical polarization, in order to align the polarization of the state to poling vector of the crystal, thus maximizing the SPDC generation efficiency. To increase compactness and mechanical robustness, the two HWPs are cemented together side by side with fast axes aligned with opposite inclinations respect to the x coordinate, then receiving the name "segmented" half wave-plate (sHWP in Fig.~\ref{fig:EPS_design}). As seen in the scheme, the state right before the ppKTP crystal is $(\ket{V}_\text{L}+\ket{V}_\text{R})/{\sqrt{2}}$, where the subscripts "L" and "R" indicate the left or right path. Each components then gets down converted, resulting in the state $(\ket{VV}_\text{R}+\ket{VV}_\text{L})/{\sqrt{2}}$. Then, sHWP$_2$ rotates the L and R components $45\degree$ in the clockwise and counter-clockwise direction, respectively, obtaining the state $(\ket{++}_\text{L}+\ket{--}_\text{R})/{\sqrt{2}}$. 
$\mathrm{SP_2}$ then implements beam recombination, resulting in the output polarization-entangled state
\begin{equation}
\label{eq:bell}
\ket{\Phi}=\frac{\ket{++}+e^{i\varphi}\ket{--}}{\sqrt{2}},
\end{equation}

\noindent where the position Hilbert space $(x,y)$ has been omitted, as it is global over the state representation.\\
The phase $\varphi$ represents the optical path difference between the two parallel trajectories through our EPS. The ability to control this phase ($\varphi$ in Eq.~\eqref{eq:bell}) is crucial for non-locality tests, in E91 QKD protocol. In fact the Clauser-Horne-Shimony-Holt (CHSH) inequality can be maximally violated only if $\varphi=0,\pi$  \cite{CHSH}. Moreover, for the BBM91 protocol it is necessary to measure the entangled state with high visibility in both H-V and D-A basis, and this is possible only if $\varphi$ is fully compensated \cite{steinlechnerSources2015, wengerowskyExperiments}.
This phase control is commonly implemented with an additional component, such as an YVO crystal (yttrium orthovanadate) or a quarter-wave plate \cite{PhysRevA.60.R773}.
In our design, however, we can modify $\varphi$ simply through a controlled tilting of the second Savart Plate (SP$_2$) along the yaw axis (indicated by angle $\uptheta$ in Fig.~\ref{fig:EPS_design}). 
This tilting of SP$_2$ induces a difference in $n_\mathrm{eff}$ between the paths of $\ket{++}$ and $\ket{--}$, resulting in a controllable relative phase.
Therefore, this enables modulating $\varphi$, granting on-demand real-time access to both the Bell states $\ket{\Phi^\pm}$, without the need for any additional external components. A geometrical explanation is given in section \ref{sec:walkof}.\\
\vspace{-0.4cm}
\vspace{-1.2cm}

\subsection{Implementation}

The two SPs (United Crystals) in our EPS are made of calcite, and produce a nominal shear of $S=\SI{1}{mm}$ at their respective working wavelengths ($\lambda_{p} = \SI{405}{\nano\meter}$ for SP$_{1}$; $\lambda_{s,i} = \SI{810}{\nano\meter}$ for SP$_{2}$).
The sHWPs (Cening, calcite on BK7 substrate) have dimensions height $\times$ width $\times$ length $ = 5 \times 10 \times 1 $ \text{mm}$^{3}$, with fast axes at $\pm22.5$. The interface between both sections of each sHWP was smaller than $\SI{10}{\upmu m}$ to minimize blocking the generated beams. However, the implementation of non-collinear generation resulted in part of the SPDC cone being lost. %, as shown in Fig. \ref{fig:FF_split}.

\begin{figure}[htb]
\begin{center}
\includegraphics[scale=0.24]{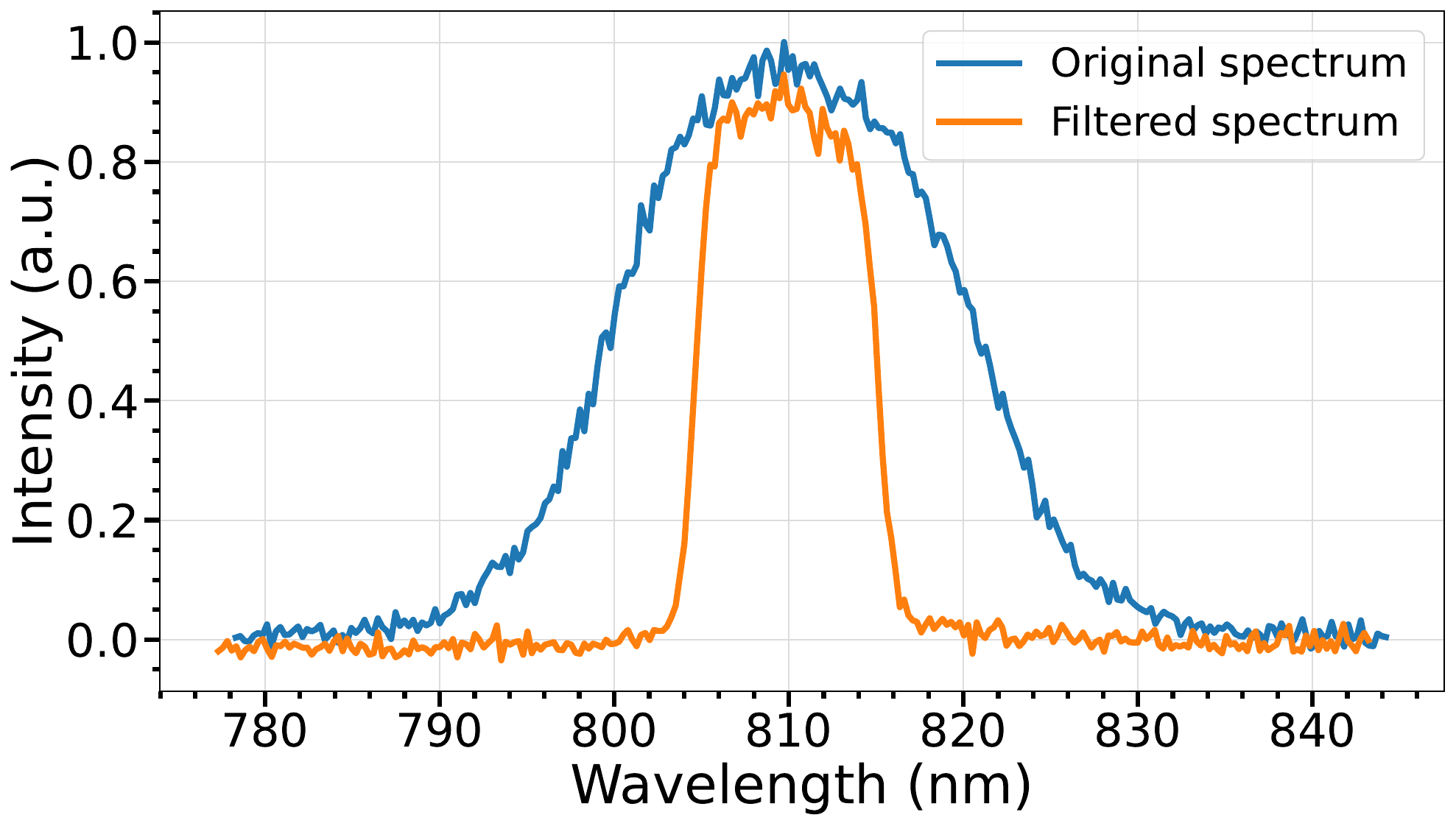}
\vspace{-1mm}
\caption{\label{fig:SPDC_spectrum} \textbf{Emission spectrum}. Generated SPDC spectrum (blue), with $\Delta\lambda=\SI{22}{nm}$ (FWHM), filtered spectrum (orange) $\Delta\lambda=\SI{10}{nm}$ (FWHM).
}
\end{center}
\end{figure}

For the nonlinear medium, we used a type-0 periodically poled potassium titanyl phosphate (ppKTP) crystal (Raicol) \cite{Comp_EPS_designs, SPDC_EPS_rev} with dimensions height $\times$ width $\times$ length $=1\times2\times20[\text{mm}^{3}]$. The pump laser is continuous-wave, single-frequency, centered at $\lambda_\mathrm{p}=\SI{405}{nm}$ (Omikron, Bluephoton series), and was focused into the ppKTP crystal by a $\SI{200}{\milli\meter}$ lens. This configuration yielded a measured $1/e^{2}$ beam waist of $2w_{0}=92\pm 2 \SI{}{\upmu \meter}$ at the crystal's center. After the second Savart plate (SP$_2$ in Fig. \ref{fig:EPS_design}) the pump light was subtracted via a band-pass and long-pass subsequent filters.
Finally, the generated SPDC photon pairs were filtered using a $810\pm \SI{5}{nm} $ band-pass filter  (see Fig.~\ref{fig:SPDC_spectrum}), in order to improve photon indistinguishability and remove background, such as fluorescence, generated in the optical components.

\vspace{-1cm}

\subsection{Performance analysis}
\label{sec:perf}

We demonstrated and quantified the source performance in two ways: performing two-photon interference, and verifying the experimental violation of CHSH inequality.

\begin{figure}[htb]
\begin{center}
\includegraphics[scale=0.24]{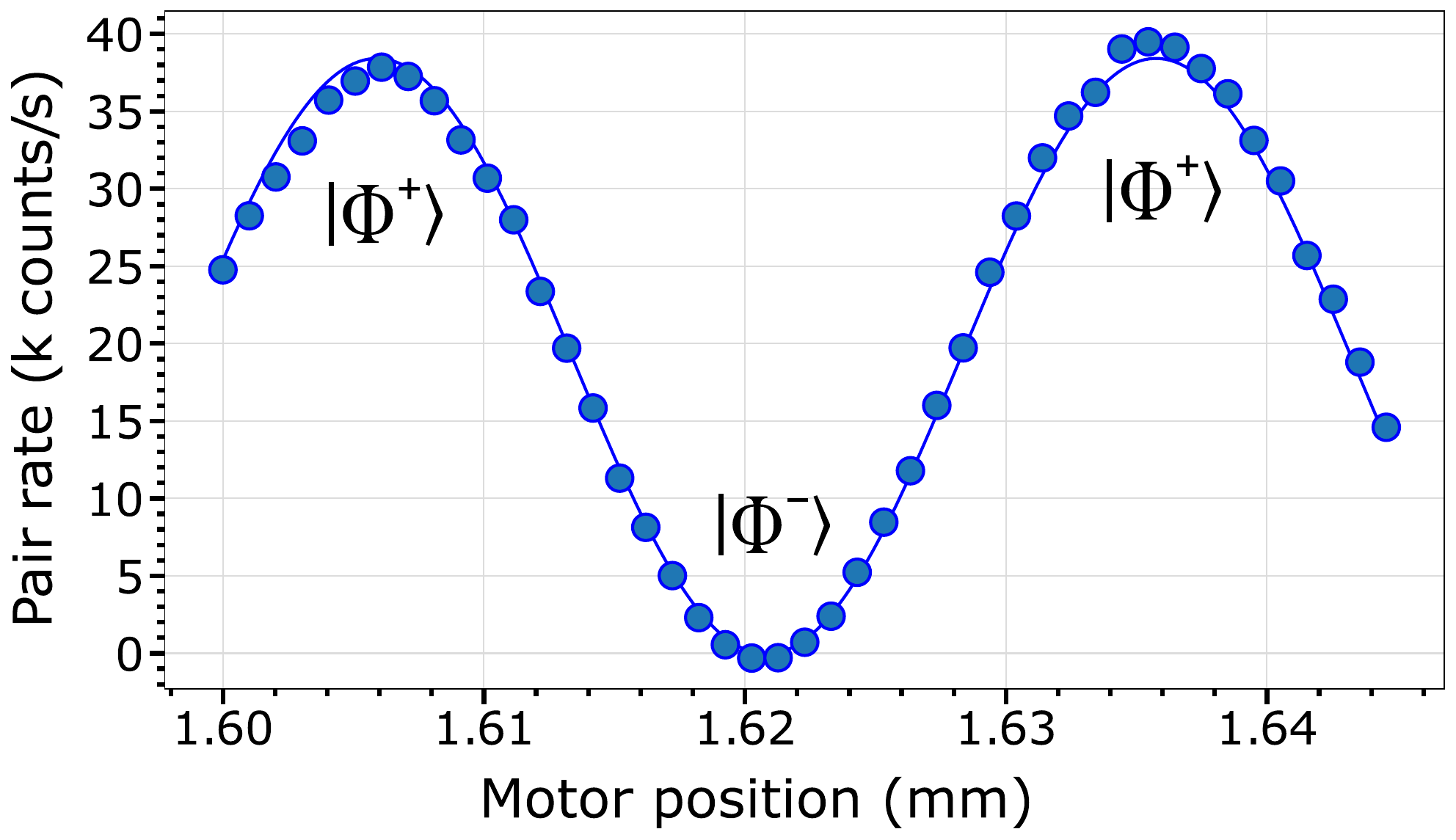}
\vspace{-1mm}
\caption{\label{fig:SP_interf} \textbf{Varying entangled state by Savart plate tilting.}
Two-photon coincidences vs.~stepper motor position; $\varphi$ in eq.~\eqref{eq:bell} is modulated by a motor tilting SP$_2$.
Blue circles, measured coincidences; solid line, cosine fitting function.
}
\end{center}
\end{figure}

Firstly, we measured two-photon interference on our source's output state in order to certify the Savart plate tilting method.
We projected the generated state $\ket{\Phi}$ (eq.~\eqref{eq:bell}) onto the $\ket{H}$ basis using a polarizer, while varying $\varphi$ in eq.~\eqref{eq:bell} by tilting SP$_2$. Figure~\ref{fig:SP_interf} shows the two-photon coincidences measured at the output, demonstrating the successful transition between the states $\ket{\Phi^+}$ and $\ket{\Phi^-}$. See Appendix \ref{sec:app0} for details.

\begin{figure}[htb]
\begin{center}
\includegraphics[scale=0.39]{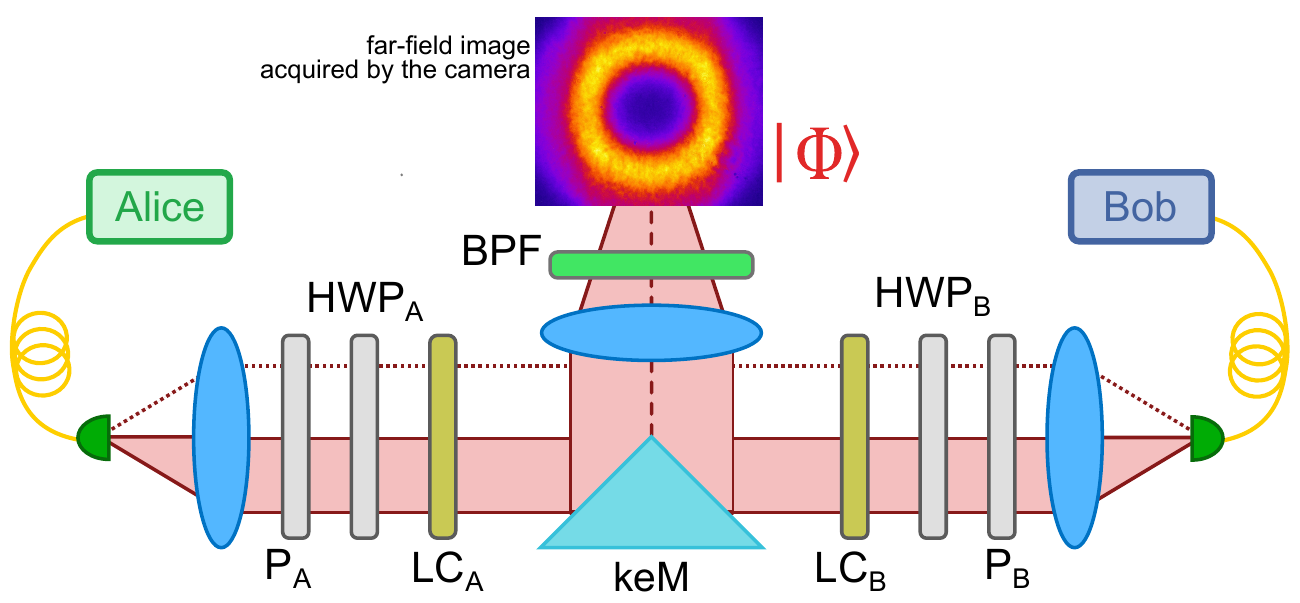}
\caption{\label{fig:FF_split} \textbf{CHSH setup}. Far-field plane of SPDC beam split in half with knife-edge mirror (keM).
Due to SPDC momentum anti-correlation, photon pairs are deterministically divided between Alice (A) and Bob (B) arms.
}
\end{center}
\vspace{-0.6cm}
\end{figure}

Secondly, we quantified the generated entanglement via violation of the CHSH inequality \cite{CHSH, Aspect}.
As shown in Fig.~\ref{fig:FF_split}, our CHSH characterization setup separates the entangled photon pairs, sending them to two independent measurement stations, Alice (A) and Bob (B).
We used a deterministic spitting method, in which a knife-edge mirror (keM) is placed in the SPDC Fourier plane of a $\SI{200}{mm}$ lens. The momentum anti-correlation of the SPDC photon pairs here corresponds to opposite positions in the transverse plane \cite{SPDC_EPS_rev}, allowing us to divide the ring-shape emission into left and right halves, which are directed to Alice and Bob, respectively.
We measure coincidences between A and B, using the liquid crystal retarders (LC$_{A}$ and LC$_{B}$ in Fig.~\ref{fig:FF_split}) and polarization projection setups (HWPs and polarizers (P) in Fig.~\ref{fig:FF_split}) to obtain the Bell curves shown in Fig.~\ref{fig:CHSH_results}. See Appendix \ref{sec:appA} for details.

\begin{figure}[htb]
\includegraphics[width=0.5\textwidth]{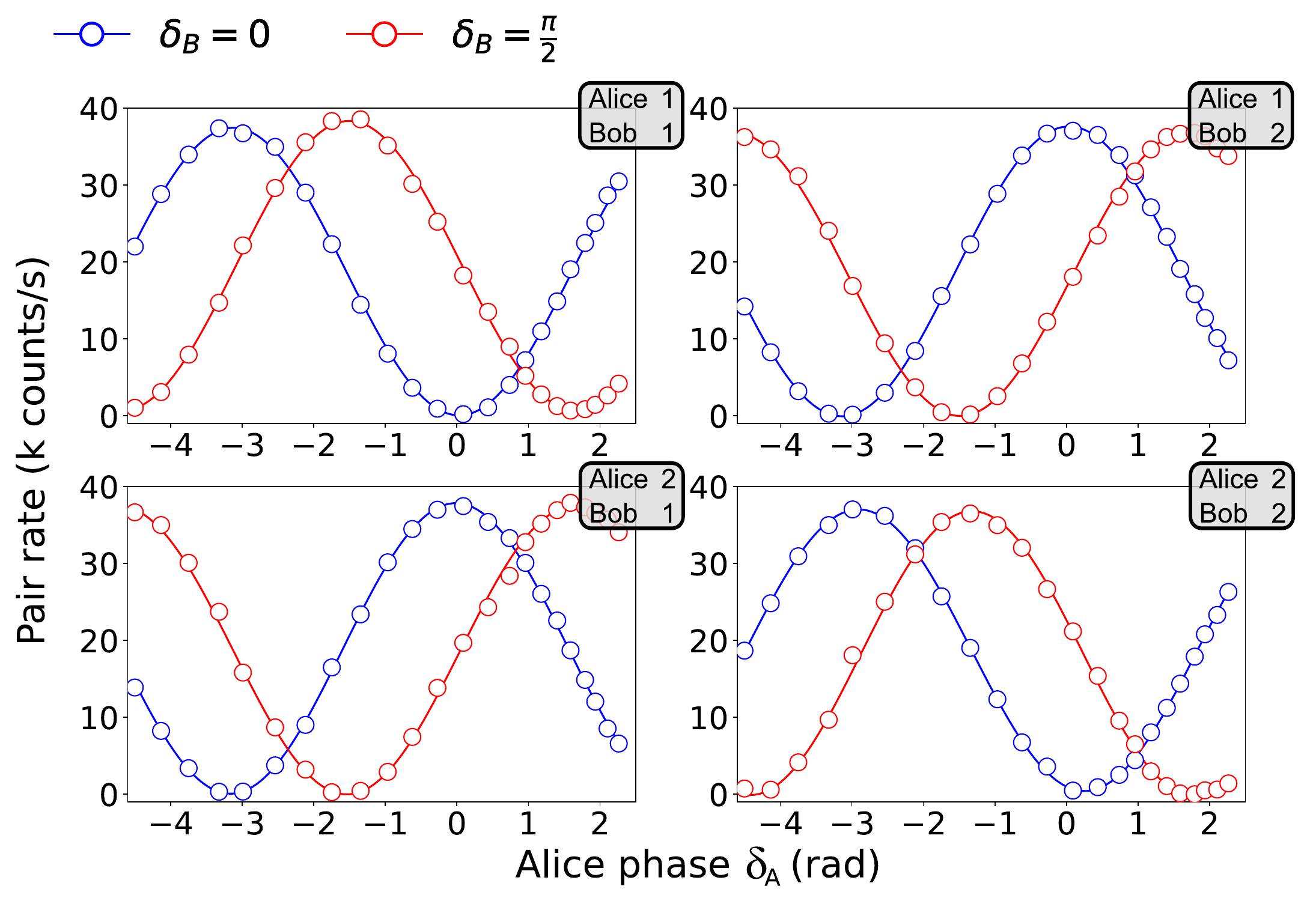}
\caption{\textbf{CHSH results} 
Coincidences between A and B, varying Alice phase $\delta_A$ for fixed Bob phase $\delta_B=0,\frac{\pi}{2}$.
%The $\delta_B$ variable refers to the fixed delay imposed to the \textit{Bob} channel.
Measurement settings 1(2) (grey boxes) correspond to horizontal(vertical) detected polarization.
Circles, measured coincidences (Poissonian standard error smaller than marker size); solid lines, fitting functions.
%The standard deviation of the detected photon count rate (Poisson errors) is smaller than the markers.
}
\label{fig:CHSH_results}
\end{figure}
We detected a total coincidence rate of $CC = (2.470\pm0.005) \times 10^5 \text{ pairs/s}$ and singles rate {$SC =2.058\pm0.003 \times 10^6 \text{ counts/s}$, at pump power $P=2.6 \;\si{\milli\watt}$, corresponding to a brightness of $(9.50\pm0.03) \times 10^4 {\text{ pairs}}/{(\text{s}\cdot \text{mW})}$ and heralding efficiency $\eta=12\%$.
We therefore estimated the emitted brightness to be $\tfrac{CC}
{\eta^2}\approx 7 \times 10^6 \, \text{pairs}/( \si{\second} \cdot \si{\milli \watt} )$.
The average fringe visibility of the fitting Bell curves (coincidences with accidentals subtracted) is $V>0.99\pm0.01$. The final CHSH parameter was computed similarly to \cite{Aspect}, obtaining $S_{\text{exp}} = 2.82 \pm 0.04$, giving a violation of the Bell inequality by 20$\sigma$.

From such $S_{\text{exp}}$ we then obtain the state Bell state visibility, using the expression $S_{\text{exp}}=\mathcal{V}_\text{S}\cdot2\sqrt{2}$. Assuming a mixed-state Werner representation $\rho=\mathcal{V}_\text{S}\ket{\Phi}\bra{\Phi}+\frac{1-V_\text{S}}{4}\mathbb{I}\otimes\mathbb{I}$, as the worst possible scenario, the two-qubit fidelity corresponds to \cite{riedelgardingBellDiagonalWerner2021}:
\begin{equation}
    \label{eq:fidelity}
    F=\frac{1}{4}{\left(\frac{3}{2}\sqrt{\mathcal{V}_\text{S}}+\frac{1}{2}\sqrt{4-3\mathcal{V}_\text{S}}\right)}^{2}.
\end{equation}
We therefore estimate a lower bound fidelity of $F=0.992\pm0.001$, with error given by Gaussian propagation.\\
A brief overview of the source performance is displayed in table \ref{table:results}.

\begin{table}
  \centering
  \begin{tabularx}{0.45\textwidth}{Xl} % X for a flexible-width column, l for left-aligned
    \toprule % Top rule
    \textbf{Specification} & \textbf{Mean value} \\
    % Bold column headers
    \midrule % Middle rule
    CC ($1/(\text{s}\cdot \text{mW})$ & $95 \times 10^3$ \\
    Heralding $ \eta $ & 12\% \\
    CAR & 14 \\
    Visibility & $99\%$ \\
    Fidelity (lower bound) & 0.99 \\
    CHSH parameter $ S $ & $2.82 \pm 0.04$ \\
    \bottomrule % Bottom rule
  \end{tabularx}
  \caption{\textbf{EPS measured performance.} CC: coincidence brightness adjusted for detection efficiency (CHSH inequality measurement), $\eta=CC/SC$: heralding efficiency, CAR: coincidences to accidentals rate}
  \label{table:results}
\end{table}

\section{Compact EPS advantages}

\subsection{SPaDEs architecture vs state-of-the-art}

The state-of-the-art for polarization-entangled photon sources is represented by the Sagnac architecture, which many papers have reported to deliver excellent results \cite{PhysRevA.73.012316, brambila, SPDC_EPS_rev}. The Sagnac indeed offers a phase-stable platform; however, it comes at the expense of wavelength-dependent performance (polarization extinction, polarization rotation, reflectivity, etc.) due to the necessity for dual-wavelength components. Moreover, the experimentally challenging alignment and the high precision required when positioning the crystal at the center of the loop drastically hinder the scalability of this design for field applications, due to limited compactness and miniaturization. Slight misalignment of the crystal position can indeed modify differently the counter-propagating SPDC generations, thus increasing both longitudinal and lateral cross-talk between photons \cite{beckert_2019}. Solving these issues is critical for paving the way both to large-scale on-field applications and the realization of commercial ready-to-use devices. The linear beam displacer based sources \cite{dual_cryst_eps, linearDisp_EPS, linear_eps} propose to solve these problems, with a simple, compact and robust platform, while maintaining a highly scalable and miniaturizable design. \\
Our architecture is an evolution of those designs, aimed at improving their mechanical sensitivity and removing the walkoff between the two polarization paths. In the following section, the walkoff criticality is analyzed, and is shown the superiority of the SPaDEs design over the single beam displacer architecture. \\
Thanks to the symmetry between the two paths $\ket{++}$ and $\ket{--}$, in fact, our EPS geometry offers great compatibility with all classes of nonlinear crystals. Differently from normal unbalanced beam displacer sources, the design can be implemented for non-uniform longitudinal poling \cite{chirped_SPDC} and/or for arbitrarily short non-linear crystals. This is necessary to ease the path towards miniaturization, for example via micro-optical platforms, where components of the EPS can be manufactured already at the right orientations and then assembled together in a single monolithic design.\\
We can also affirm that our source is at least equally mechanically stable to an analogous Beam displacer source (see Supplement 1).

\subsection{Longitudinal Walkoff}
\label{sec:walkof}

As explained in the previous section, the Savart plate-based design ensures that the two SPDC generation modes focus in the exact same longitudinal position in the nonlinear crystal.
In order to demonstrate the impact of non-balanced beam displacing in the generation, superposition and collection of entangled photons, we reproduced the case in which a focused beam passes through a BD element, as displayed in Fig. \ref{fig:walkoff_a}. We measured the relative longitudinal locations of the two beam waists after the BD. We performed the measurement at $\lambda=\SI{405}{\nano m}$,  for a BD with a lateral shear of $S_\mathrm{BD}=1.010 \pm 0.001 \,\si{\milli \metre}$ and for a Savart Plate with $S_\mathrm{SP}=0.972 \pm 0.001\, \si{\milli \metre}$. The light source, were initially collimated to a diameter $D=\SI{1.4}{mm}$, and focused down to $2w_{0}\sim\SI{20}{\upmu \metre}$ by a $f_1=\SI{150}{mm}$ lens.

The beam displacer (and Savart Plate) were placed in the middle point between the $f_1$ lens and $w_{0}$, simulating the EPS configuration. We detected the beam profile using a CMOS camera, that was longitudinally translated along the optical $z$-axis by means of a motorized stage, in order to capture the different focusing planes. The measured beam waists in each $z$ position is reported in Fig. \ref{fig:walkoff_b}, for both the single beam displacer and for the Savart plate.

\begin{figure}[h]
    \centering
    \subfigure[]{\captionsetup{justification=raggedright,singlelinecheck=false}\includegraphics[width=0.45\textwidth]{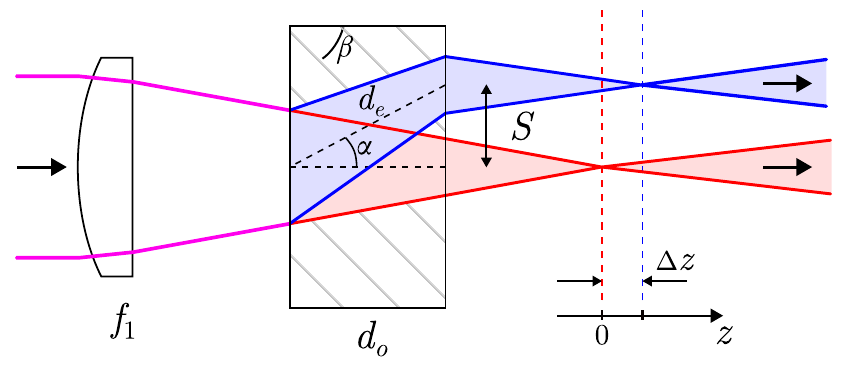}\label{fig:walkoff_a}} 
    \subfigure[]{\captionsetup{justification=raggedright,singlelinecheck=false}\includegraphics[scale=0.24]{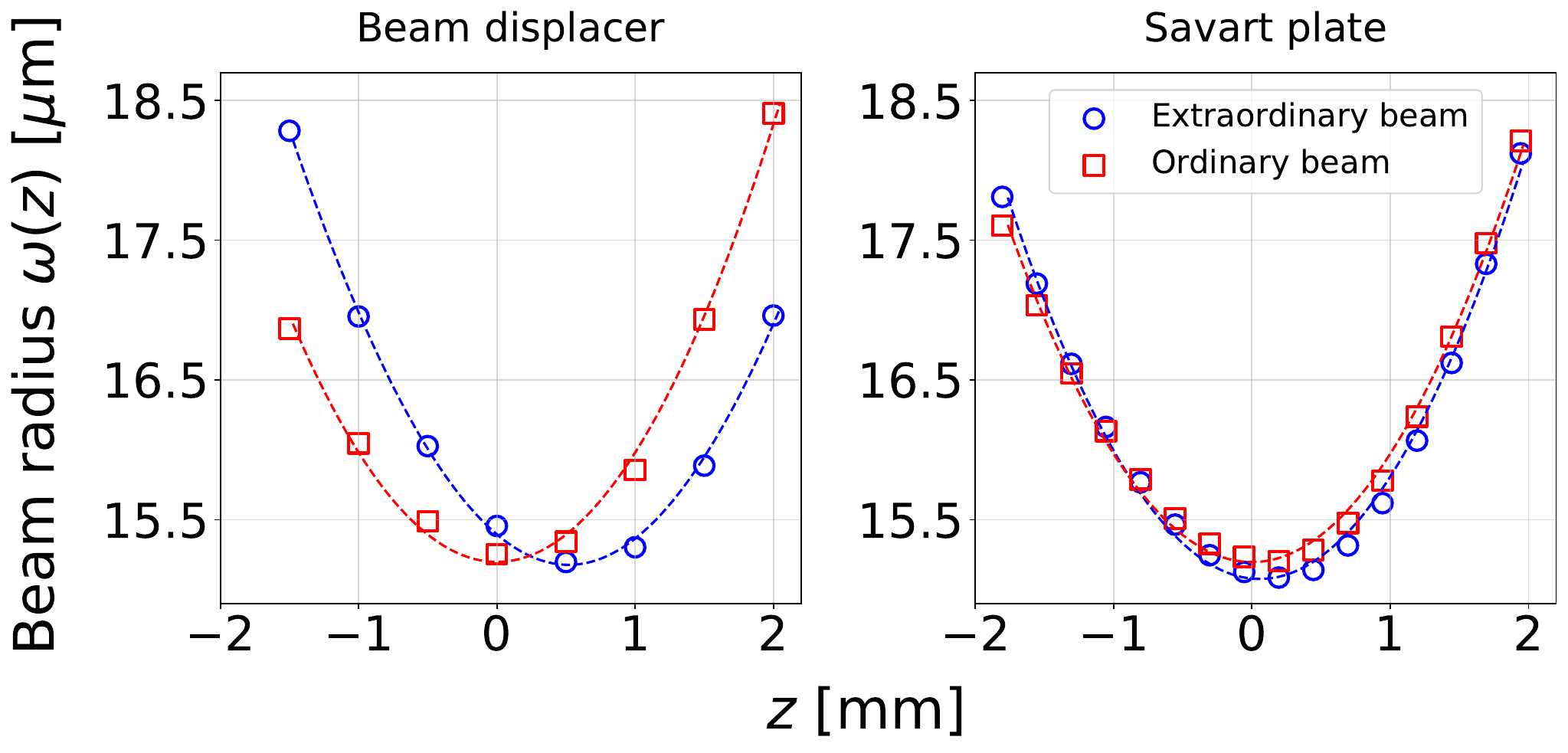}\label{fig:walkoff_b}} 
    \caption{(a) Diagram of longitudinal walkoff in a BD element using a focused beam. Vertical segmented lines at the valleys of the hyperbolas indicating associated value (b) Results for UV and NIR cases.} \label{fig:walkoff}
\end{figure}

For the BD element with $d_\text{o}=\SI{8.73}{mm}$, the expected walkoff is $\Delta z_\mathrm{BD}=\SI{0.542}{mm}$ (see Appendix \ref{sec:appB}). The experimental result, extrapolated from the fitting functions in Fig. \ref{fig:walkoff_b}, is  $\Delta \bar{z}_\mathrm{BD}=0.52 \pm 0.03 \, \si{\milli \metre}$. For the Savart plate we obtained a walkoff of $\Delta \bar{z}_\mathrm{SP}=0.06 \pm 0.03 \, \si{\milli \metre}$. We see that $\Delta \bar{z}=0.52_\mathrm{BD}$ is at least one order of magnitude higher than standard crystal periodicities, making the BD an unsuitable component when utilizing chirped poling periods. This also limits the minimum length of the crystal one can employ: for $\sim 1mm$ long crystals, by placing the waist of one polarization in the center of it, we already would find the waist of the orthogonal component outside the crystal. 
The Savart plate walkoff did not appear to be perfectly null due to experimental error. Also some minor astigmatism typical of these components could have affected the accuracy of this measurement.

\section{Discussion}

The estimated brightness of our source was $CC = (9.50\pm0.03) \times 10^4 {\text{ pairs}}/{(\text{s}\cdot \text{mW})}$, which is sufficient for achieving optimal secure key rates in realistic free-space or fiber entanglement-based QKD applications \cite{eckerStrategiesAchievingHigh2021a}. Our source brightness (table \ref{table:results}) is, however, lower than the maximum reported values in the literature \cite{brambila}. We attribute this mainly to the choice of an SPDC working-point in the non-collinear phase-matching regime, which induces a lower spectral brightness compared to the collinear case \cite{steinlechnerEfficientHeraldingPolarizationentangled2014b}.
Nonetheless, the advantage of the non-collinear regime is that it enables deterministically splitting the generated photon pairs, without relying on non-degenerate spectral correlations, as would be required when using a sharp dichroic mirror or wavelength-division multiplexing devices.

Optical losses in our EPS are attributed to the optical components' transmissivities ($78\%$) and fiber coupling. In particular, sHWP$_2$ (see Fig.~\ref{fig:EPS_design}), at the interface between the two segmented-HWP segments, partially absorbs a portion of each SPDC far-field ring ($\sim10\%)$. Moreover, the knife-edge mirror (keM in Fig. \ref{fig:FF_split}), due to fabrication limits, induces some diffraction and scattering effects that cause additional losses ($\sim 15\%$) to light in proximity of the edge.
%On the other hand, the choice of adopting a non-collinear temperature regime was driven by the choice of a deterministic pair splitting approach that would not reside in the necessity of a sharp dichroic mirror or wavelenght-division multiplexers.

We note that, when seeking compactness and high scalability, it is natural to mention integrated platforms, which offer some advantages over traditional bulk sources. In particular, waveguide-based sources excel in brightness due to their intense light confinement. These can be specifically engineered to produce light with desirable single-mode characteristics (spatially and spectrally) \cite{waveg_OPO_ent, waveguide_entang}.
%Alternatives like four-wavemixing \cite{four_wave, four_wave_mix}, implemented in photonic crystal fibers and waveguides, represents another approach. However, due to the third-order nature of this nonlinear process, they fall short in achieving the brightness of bulk-based systems.
Photonic Integrated Circuits (PICs) mark a significant advancement for vertically integrating quantum technologies \cite{silicon_PIC, silicon_PIC2}, given their scalable manufacturing processes.
%Despite these advancements, the current state of integrated systems does not yet match the entanglement quality achievable with bulk systems.
However, at present the versatility and modularity of such PIC systems are limited, as they cannot harness multi-modal characteristics like position, momentum, and orbital angular momentum correlations.
Additionally, integration is feasible only for narrow-band, single-mode systems, and efficiently out-coupling from PIC EPSs to free-space or fibre remains challenging \cite{SPDC_EPS_rev}.
%highlighting the gap that still exists in comparison to the more mature bulk architectures.
Lastly, a promising candidate for the next generation of quantum technologies are semiconductor-based quantum dots that provide on-demand entangled photon pairs with narrow linewidth and negligible multiphoton events \cite{qdots_3, qdots_rev}. However, quantum dot sources have not yet attained the performance levels of their bulk EPS counterparts, both in terms of brightness, and in state fidelity \cite{exciton_fidelity, qdots, qdots_2}, and typically require cryogenic temperatures that limit field deployment and scalability.

\section{Conclusion}
In this work we have shown a SPDC-based polarization entangled photon pair source (EPS) design, based on the evolution of previous linear beam-displacers architectures. This design is based on laterally displacing the $\ket{+}$ and $\ket{-}$ components of a vertically polarized pump beam, and sending these parallel modes through the nonlinear crystal. The two generated SPDC modes are then recombined to obtain the bell state $\ket{\Phi}=\frac{\ket{++}+e^{i\varphi}\ket{--}}{\sqrt{2}}$. The parallel splitting of the pump and recombination of the SPDC modes is performed by a pair of Savart plates, that symmetrically displace each polarization mode. This is an evolution serves the purpose to completely remove the the longitudinal walkoff from inside the crystal.

Our EPS generates polarization entangled photons, in the form of N00N states or Bell states, easily tunable between the $\ket{\Phi^+}$ and $\ket{\Phi^-}$ states through motorized tilting of the second Savart plates. Characterizing the entanglement generated by our source, we measured a CHSH parameter of $ S = 2.82\pm0.04$ ($\sim20 \sigma$ violation of the CHSH inequality), which implies a near-ideal fidelity of $F=0.992\pm0.001$, from \cite{fidelity_visibility}. The brightness of our EPS was measured to be $CC = (9.50\pm0.03) \times 10^4 {\text{ pairs}}/{(\text{s}\cdot \text{mW})}$, which gives an estimated entangled photon pair emission rate of $CC_{\mathrm{em}}\! = \! 105 \times 10^6 \, \text{pairs}/( \si{\second} \cdot \si{\milli \watt} )$ when taking into account system losses. We also studied the mechanical robustness, and longitudinal walkoff of our EPS. In comparison to existing standard BD designs, we found that our source drastically reduces longitudinal walkoff, while maintaining near-identical mechanical stability.

Our design combines a series of unique ideas, whose implementation offers high performance either for i) scalability: all components can be easily miniaturized as long as the two beams pass through independent sections of the sHWPs, which is compatible even with beams that cross each other at the center of the crystal due to additional focusing elements; ii) harsh environments: bulk optics offers one of the best generation stabilities under thermal/pressure fluctuations, making our EPS an ideal candidate for space quantum communications.

%Natural future upgrades are represented by the employment of superior segmented half-wave plates, in order to remove the heralding losses at each step of the source, as well as the development of an engineering model that could miniaturize the setup and enable real-environment qualification. An obvious follow-up would also be a campaign of tests to reach space qualification of the design, given its highly desirable characteristics.

Future upgrades will include superior segmented half-wave plates in order to improve the heralding efficiency, as well as development of an engineering model to further miniaturize the setup and enable real-environment qualification, such as space qualification.

%%%%%%%%%%%%%%%%%%%%%%%%%%%%%%%%%%%%%%%%%%%%%%%%%%%%%%%%%%%%%
%\begin{comment}
\section*{Acknowledgments}
This study was supported by MCIN with funding from European Union NextGenerationEU(PRTR-C17.I1) and by Generalitat de Catalunya.
This project has received funding from the European Union’s Horizon Europe research and innovation programme under the project "Quantum Security Networks Partnership" (QSNP, grant agreement No 101114043).
This work was partially funded by CEX2019-000910-S [MCIN/ AEI/10.13039/501100011033], Fundació Cellex, Fundació Mir-Puig, and Generalitat de Catalunya through CERCA.\\
G. P. acknowledges funding from the European Union's Horizon Europe research and innovation programme under the Marie Skłodowska-Curie Grant Agreement No 101081441. A.C. work was Funded by CEX2019-000910-S [MCIN/ AEI/10.13039/501100011033], Fundació Cellex, Fundació Mir-Puig, and Generalitat de Catalunya through CERCA.
R. C. acknowledges funding from the European Union's Horizon Europe research and innovation programme under the Marie Sklodowska-Curie Grant Agreement No 713729 (ICFOstepstone 2), and from the European Union’s Horizon Europe research and innovation programme under Grant Agreement No 101082596 (QUDICE).
A.D. acknowledges support of the ICFO CELLEX PhD-fellowship. 
%\end{comment}

\section*{Disclosures}
AC, RC, AD, and VP are co-inventors of a patent application related to the content of this paper.

\appendix
%%%%%%%%%%%%%%%%%%%%%%%%%%%%%%%%%%%%%%%%%%%%%%%%%%%%%%%%%%%%%
\section{Savart-plate tilting phase-shifting}
\label{sec:app0}
To demonstrate that tilting the Savart plate SP$_2$ allows tuning $\varphi$ in eq.~\ref{eq:bell}, we project the generated state $\ket{\Phi^+}$ onto H using a polarizer. This output is single-mode fibre (SMF) coupled, and the pairs are split probabilistically with a 50-50 fibre beam-splitter (FBS, Thorlabs).
We measure two-photon coincidences between the FBS outputs $a$ and $b$, using a pair of single-photon avalanche diode (SPAD) detectors, extracting coincidences using a $ \SI{1}{ns} $ window, and subtracting accidentals.
These coincidences depend on $\varphi$ in eq.~\eqref{eq:bell} as:
\begin{equation}
\label{eq:interf}
P_{a,b}\equiv|\braket{H_a,H_b|\Phi}_\text{FBS}|^{2}=\frac{1+\mathcal{V}_\text{SP}\cos(\varphi)}{2},
%P_{a,b}\equiv|\braket{0|a_{H}b_{H}|\Phi}_\text{FBS}|^{2}=\frac{1+\mathcal{V}_\text{SP}\cos(\varphi)}{2},
\end{equation}
where we have introduced the interference visibility ($V_\text{SP}$). This coincidence curve is shown in Fig.~\ref{fig:SP_interf}, where we vary the tilting angle $\uptheta$ of SP$_2$ with respect to the optical axis (see Fig.~\ref{fig:EPS_design}).

SP$_2$ tilting is implemented using a stepper motor pushing one side of SP$_2$ along the z-axis.
%Since we are translating over small distances ($\sim 1 \si{\micro\meter}$) we approximate $\varphi\propto\sin(\theta)\propto z$. Therefore, the acquired coincidence curve from equation \eqref{eq:interf} is reported in Fig. \ref{fig:SP_interf}, as (coinc. rate vs $z$), where $z$ is the step motor relative position. The coincidences are corrected from accidentals contribution. \\
%The fitting confirmed the working principle with a confidence level of $p\!>\!99.99\%$.
The coincidence interference visibilities with and without accidentals subtraction were $\mathcal{V}_\text{sp} = 1.000 \pm 0.010$ and $\widetilde{\mathcal{V}}_\text{sp} = 0.840 \pm 0.008$, respectively. This result also shows that SMF-coupling remains unaffected by the Savart plate tilting.\\

%%%%%%%%%%%%%%%%%%%%%%%%%%%%%%%%%%%%%%%%%%%%%%%%%%%%%%%%%%%%%
\section{CHSH inequality}
\label{sec:appA}
As shown in Fig.~\ref{fig:FF_split}, Alice (Bob) side contains a liquid crystal LC$_{A}$ (LC$_{B}$) placed at 45-degrees rotation, to implement $\varphi_{A}$ ($\varphi_{B}$) phase between vertical and horizontal polarizations.
The half-wave plates (HWP${A}$ and HWP${B}$) and polarizers (P${A}$ and P${B}$) in each arm enable the projection into vertical and horizontal polarizations. By alternating between these polarization states, we can measure coincidences across various polarization combinations without needing the four detectors typically required in Aspect-type experiments\cite{Aspect}. This is achieved by simply rotating the polarization of the state.
The light in A and B is coupled into SMFs, and sent to SPADs for coincidence detection ($\SI{1}{ns} $ window), with accidentals removed.
The optimal Bell angles for Bob are $\tfrac{\pi}{2}$ apart \cite{Bell}.
Therefore, the liquid crystal LC$_\text{B}$ in Fig. \ref{fig:FF_split} was fixed to $\delta_B=0,\frac{\pi}{2}$, while LC$_\text{A}$ was scanned over an entire period, in order to find the $\delta_A$ values which maximize the inequality.
For each Bob phase setting ($\delta_B=0,\frac{\pi}{2}$) we obtained four coincidence curves for four combinations of the four channels Alice 1,2 and Bob 1,2. Here Alice 1(2), and Bob 1(2)  correspond to horizontal (vertical detected polarization. These results are shown in Fig. \ref{fig:CHSH_results}.

%%%%%%%%%%%%%%%%%%%%%%%%%%%%%%%%%%%%%%%%%%%%%%%%%%%%%%%%%%%%%
\section{Longitudinal walkoff measurement}
\label{sec:appB}

As shown in the optical scheme of Fig.\ref{fig:walkoff}(a), the extraordinary beam path corresponds to $d_\text{e}=\sqrt{d_\text{o}^{2}+S^{2}}$, with $d_\text{o}$ as the ordinary beam path that also coincides with the BD width. The optical paths difference can be calculated as 
\begin{equation}
    \Delta z = d_\text{o}n_\text{o}-d_\text{e}n_\text{e},
\end{equation}
\noindent where 
\begin{equation}
\label{eq:refr}
    n=1/\sqrt{\frac{\cos^{2}(\alpha-\beta)}{n_\text{o}^{2}}+\frac{\sin^{2}(\alpha-\beta)}{n_\text{e}^{2}}}
\end{equation}
is the effective refractive index for the non-ordinary polarized light. Here $\alpha=\arctan(S/d_\text{o})$ is the deviation angle and $\beta$ the optical axis angle. The refractive indexes of the calcite for ordinary (extraordinary) polarized light is $n_\text{o}=1.66$ ($n_\text{e}=1.49$).
Our BDs were manufactured with $\beta=\pi/4$: the UV element with $d_\text{o}=\SI{8.73}{mm}$ and the NIR one with $d_\text{o}=\SI{11.26}{mm}$. Assuming that both focal points are approximately equally displaced by the prismatic effect of the BD, we predict $\Delta z_\text{UV}=\SI{0.542}{mm}$ $\Delta z_\text{NIR}=\SI{0.73}{mm}$.

\bibliography{mybib}

%merlin.mbs aipnum4-1.bst 2010-07-25 4.21a (PWD, AO, DPC) hacked
%Control: key (0)
%Control: author (8) initials jnrlst
%Control: editor formatted (1) identically to author
%Control: production of article title (0) allowed
%Control: page (1) range
%Control: year (1) truncated
%Control: production of eprint (0) enabled
\begin{thebibliography}{50}%
\makeatletter
\providecommand \@ifxundefined [1]{%
 \@ifx{#1\undefined}
}%
\providecommand \@ifnum [1]{%
 \ifnum #1\expandafter \@firstoftwo
 \else \expandafter \@secondoftwo
 \fi
}%
\providecommand \@ifx [1]{%
 \ifx #1\expandafter \@firstoftwo
 \else \expandafter \@secondoftwo
 \fi
}%
\providecommand \natexlab [1]{#1}%
\providecommand \enquote  [1]{``#1''}%
\providecommand \bibnamefont  [1]{#1}%
\providecommand \bibfnamefont [1]{#1}%
\providecommand \citenamefont [1]{#1}%
\providecommand \href@noop [0]{\@secondoftwo}%
\providecommand \href [0]{\begingroup \@sanitize@url \@href}%
\providecommand \@href[1]{\@@startlink{#1}\@@href}%
\providecommand \@@href[1]{\endgroup#1\@@endlink}%
\providecommand \@sanitize@url [0]{\catcode `\\12\catcode `\$12\catcode `\&12\catcode `\#12\catcode `\^12\catcode `\_12\catcode `\%12\relax}%
\providecommand \@@startlink[1]{}%
\providecommand \@@endlink[0]{}%
\providecommand \url  [0]{\begingroup\@sanitize@url \@url }%
\providecommand \@url [1]{\endgroup\@href {#1}{\urlprefix }}%
\providecommand \urlprefix  [0]{URL }%
\providecommand \Eprint [0]{\href }%
\providecommand \doibase [0]{http://dx.doi.org/}%
\providecommand \selectlanguage [0]{\@gobble}%
\providecommand \bibinfo  [0]{\@secondoftwo}%
\providecommand \bibfield  [0]{\@secondoftwo}%
\providecommand \translation [1]{[#1]}%
\providecommand \BibitemOpen [0]{}%
\providecommand \bibitemStop [0]{}%
\providecommand \bibitemNoStop [0]{.\EOS\space}%
\providecommand \EOS [0]{\spacefactor3000\relax}%
\providecommand \BibitemShut  [1]{\csname bibitem#1\endcsname}%
\let\auto@bib@innerbib\@empty
%</preamble>
\bibitem [{\citenamefont {Gisin}\ \emph {et~al.}(2002)\citenamefont {Gisin}, \citenamefont {Ribordy}, \citenamefont {Tittel},\ and\ \citenamefont {Zbinden}}]{GisinRev}%
  \BibitemOpen
  \bibfield  {author} {\bibinfo {author} {\bibfnamefont {N.}~\bibnamefont {Gisin}}, \bibinfo {author} {\bibfnamefont {G.}~\bibnamefont {Ribordy}}, \bibinfo {author} {\bibfnamefont {W.}~\bibnamefont {Tittel}}, \ and\ \bibinfo {author} {\bibfnamefont {H.}~\bibnamefont {Zbinden}},\ }\bibfield  {title} {\enquote {\bibinfo {title} {Quantum cryptography},}\ }\href {\doibase 10.1103/RevModPhys.74.145} {\bibfield  {journal} {\bibinfo  {journal} {Rev. Mod. Phys.}\ }\textbf {\bibinfo {volume} {74}},\ \bibinfo {pages} {145--195} (\bibinfo {year} {2002})}\BibitemShut {NoStop}%
\bibitem [{\citenamefont {Kimble}(2008)}]{kimbleQuantumInternet2008}%
  \BibitemOpen
  \bibfield  {author} {\bibinfo {author} {\bibfnamefont {H.~J.}\ \bibnamefont {Kimble}},\ }\bibfield  {title} {\enquote {\bibinfo {title} {The quantum internet},}\ }\href {\doibase 10.1038/nature07127} {\bibfield  {journal} {\bibinfo  {journal} {Nature}\ }\textbf {\bibinfo {volume} {453}},\ \bibinfo {pages} {1023--1030} (\bibinfo {year} {2008})}\BibitemShut {NoStop}%
\bibitem [{\citenamefont {Bennett}\ \emph {et~al.}(1993)\citenamefont {Bennett}, \citenamefont {Brassard}, \citenamefont {Cr\'epeau}, \citenamefont {Jozsa}, \citenamefont {Peres},\ and\ \citenamefont {Wootters}}]{telep}%
  \BibitemOpen
  \bibfield  {author} {\bibinfo {author} {\bibfnamefont {C.~H.}\ \bibnamefont {Bennett}}, \bibinfo {author} {\bibfnamefont {G.}~\bibnamefont {Brassard}}, \bibinfo {author} {\bibfnamefont {C.}~\bibnamefont {Cr\'epeau}}, \bibinfo {author} {\bibfnamefont {R.}~\bibnamefont {Jozsa}}, \bibinfo {author} {\bibfnamefont {A.}~\bibnamefont {Peres}}, \ and\ \bibinfo {author} {\bibfnamefont {W.~K.}\ \bibnamefont {Wootters}},\ }\href {\doibase 10.1103/PhysRevLett.70.1895} {\bibfield  {journal} {\bibinfo  {journal} {Phys. Rev. Lett.}\ }\textbf {\bibinfo {volume} {70}},\ \bibinfo {pages} {1895--1899} (\bibinfo {year} {1993})}\BibitemShut {NoStop}%
\bibitem [{\citenamefont {Bouwmeester}\ \emph {et~al.}(1997)\citenamefont {Bouwmeester}, \citenamefont {Pan}, \citenamefont {Mattle}, \citenamefont {Eibl}, \citenamefont {Weinfurter},\ and\ \citenamefont {Zeilinger}}]{telep_zeil}%
  \BibitemOpen
  \bibfield  {author} {\bibinfo {author} {\bibfnamefont {D.}~\bibnamefont {Bouwmeester}}, \bibinfo {author} {\bibfnamefont {J.-W.}\ \bibnamefont {Pan}}, \bibinfo {author} {\bibfnamefont {K.}~\bibnamefont {Mattle}}, \bibinfo {author} {\bibfnamefont {M.}~\bibnamefont {Eibl}}, \bibinfo {author} {\bibfnamefont {H.}~\bibnamefont {Weinfurter}}, \ and\ \bibinfo {author} {\bibfnamefont {A.}~\bibnamefont {Zeilinger}},\ }\href {\doibase 10.1038/37539} {\bibfield  {journal} {\bibinfo  {journal} {Nature}\ }\textbf {\bibinfo {volume} {390}},\ \bibinfo {pages} {575–--579} (\bibinfo {year} {1997})}\BibitemShut {NoStop}%
\bibitem [{\citenamefont {Degen}, \citenamefont {Reinhard},\ and\ \citenamefont {Cappellaro}(2017)}]{sensing_rev}%
  \BibitemOpen
  \bibfield  {author} {\bibinfo {author} {\bibfnamefont {C.~L.}\ \bibnamefont {Degen}}, \bibinfo {author} {\bibfnamefont {F.}~\bibnamefont {Reinhard}}, \ and\ \bibinfo {author} {\bibfnamefont {P.}~\bibnamefont {Cappellaro}},\ }\bibfield  {title} {\enquote {\bibinfo {title} {Quantum sensing},}\ }\href {\doibase 10.1103/RevModPhys.89.035002} {\bibfield  {journal} {\bibinfo  {journal} {Rev. Mod. Phys.}\ }\textbf {\bibinfo {volume} {89}},\ \bibinfo {pages} {035002} (\bibinfo {year} {2017})}\BibitemShut {NoStop}%
\bibitem [{\citenamefont {Gilaberte~Basset}\ \emph {et~al.}(2019)\citenamefont {Gilaberte~Basset}, \citenamefont {Setzpfandt}, \citenamefont {Steinlechner}, \citenamefont {Beckert}, \citenamefont {Pertsch},\ and\ \citenamefont {Gräfe}}]{imaging_rev}%
  \BibitemOpen
  \bibfield  {author} {\bibinfo {author} {\bibfnamefont {M.}~\bibnamefont {Gilaberte~Basset}}, \bibinfo {author} {\bibfnamefont {F.}~\bibnamefont {Setzpfandt}}, \bibinfo {author} {\bibfnamefont {F.}~\bibnamefont {Steinlechner}}, \bibinfo {author} {\bibfnamefont {E.}~\bibnamefont {Beckert}}, \bibinfo {author} {\bibfnamefont {T.}~\bibnamefont {Pertsch}}, \ and\ \bibinfo {author} {\bibfnamefont {M.}~\bibnamefont {Gräfe}},\ }\bibfield  {title} {\enquote {\bibinfo {title} {Perspectives for applications of quantum imaging},}\ }\href {\doibase https://doi.org/10.1002/lpor.201900097} {\bibfield  {journal} {\bibinfo  {journal} {Laser \& Photonics Reviews}\ }\textbf {\bibinfo {volume} {13}},\ \bibinfo {pages} {1900097} (\bibinfo {year} {2019})}\BibitemShut {NoStop}%
\bibitem [{\citenamefont {Slussarenko}\ and\ \citenamefont {Pryde}(2019)}]{QuComp1}%
  \BibitemOpen
  \bibfield  {author} {\bibinfo {author} {\bibfnamefont {S.}~\bibnamefont {Slussarenko}}\ and\ \bibinfo {author} {\bibfnamefont {G.~J.}\ \bibnamefont {Pryde}},\ }\bibfield  {title} {\enquote {\bibinfo {title} {{Photonic quantum information processing: A concise review}},}\ }\href {\doibase 10.1063/1.5115814} {\bibfield  {journal} {\bibinfo  {journal} {Applied Physics Reviews}\ }\textbf {\bibinfo {volume} {6}},\ \bibinfo {pages} {041303} (\bibinfo {year} {2019})}\BibitemShut {NoStop}%
\bibitem [{\citenamefont {Madsen}\ \emph {et~al.}(2022)\citenamefont {Madsen}, \citenamefont {Laudenbach}, \citenamefont {Askarani}, \citenamefont {Rortais}, \citenamefont {Vincent}, \citenamefont {Bulmer}, \citenamefont {Miatto}, \citenamefont {Neuhaus}, \citenamefont {Helt}, \citenamefont {Collins}, \citenamefont {Lita}, \citenamefont {Gerrits}, \citenamefont {Nam}, \citenamefont {Vaidya}, \citenamefont {Menotti}, \citenamefont {Dhand}, \citenamefont {Vernon}, \citenamefont {Quesada},\ and\ \citenamefont {Lavoie}}]{QuComp2}%
  \BibitemOpen
  \bibfield  {author} {\bibinfo {author} {\bibfnamefont {L.~S.}\ \bibnamefont {Madsen}}, \bibinfo {author} {\bibfnamefont {F.}~\bibnamefont {Laudenbach}}, \bibinfo {author} {\bibfnamefont {M.~F.}\ \bibnamefont {Askarani}}, \bibinfo {author} {\bibfnamefont {F.}~\bibnamefont {Rortais}}, \bibinfo {author} {\bibfnamefont {T.}~\bibnamefont {Vincent}}, \bibinfo {author} {\bibfnamefont {J.~F.~F.}\ \bibnamefont {Bulmer}}, \bibinfo {author} {\bibfnamefont {F.~M.}\ \bibnamefont {Miatto}}, \bibinfo {author} {\bibfnamefont {L.}~\bibnamefont {Neuhaus}}, \bibinfo {author} {\bibfnamefont {L.~G.}\ \bibnamefont {Helt}}, \bibinfo {author} {\bibfnamefont {M.~J.}\ \bibnamefont {Collins}}, \bibinfo {author} {\bibfnamefont {A.~E.}\ \bibnamefont {Lita}}, \bibinfo {author} {\bibfnamefont {T.}~\bibnamefont {Gerrits}}, \bibinfo {author} {\bibfnamefont {S.~W.}\ \bibnamefont {Nam}}, \bibinfo {author} {\bibfnamefont {V.~D.}\ \bibnamefont {Vaidya}}, \bibinfo {author} {\bibfnamefont {M.}~\bibnamefont {Menotti}}, \bibinfo {author}
  {\bibfnamefont {I.}~\bibnamefont {Dhand}}, \bibinfo {author} {\bibfnamefont {Z.}~\bibnamefont {Vernon}}, \bibinfo {author} {\bibfnamefont {N.}~\bibnamefont {Quesada}}, \ and\ \bibinfo {author} {\bibfnamefont {J.}~\bibnamefont {Lavoie}},\ }\bibfield  {title} {\enquote {\bibinfo {title} {Quantum computational advantage with a programmable photonic processor},}\ }\href {\doibase 10.1038/s41586-022-04725-x} {\bibfield  {journal} {\bibinfo  {journal} {Nature}\ }\textbf {\bibinfo {volume} {606}},\ \bibinfo {pages} {75--81} (\bibinfo {year} {2022})}\BibitemShut {NoStop}%
\bibitem [{\citenamefont {Lu}\ \emph {et~al.}(2022)\citenamefont {Lu}, \citenamefont {Cao}, \citenamefont {Peng},\ and\ \citenamefont {Pan}}]{micius_rev}%
  \BibitemOpen
  \bibfield  {author} {\bibinfo {author} {\bibfnamefont {C.-Y.}\ \bibnamefont {Lu}}, \bibinfo {author} {\bibfnamefont {Y.}~\bibnamefont {Cao}}, \bibinfo {author} {\bibfnamefont {C.-Z.}\ \bibnamefont {Peng}}, \ and\ \bibinfo {author} {\bibfnamefont {J.-W.}\ \bibnamefont {Pan}},\ }\bibfield  {title} {\enquote {\bibinfo {title} {Micius quantum experiments in space},}\ }\href {\doibase 10.1103/RevModPhys.94.035001} {\bibfield  {journal} {\bibinfo  {journal} {Reviews of Modern Physics}\ }\textbf {\bibinfo {volume} {94}},\ \bibinfo {pages} {035001} (\bibinfo {year} {2022})}\BibitemShut {NoStop}%
\bibitem [{\citenamefont {Bernhard}\ \emph {et~al.}(2013)\citenamefont {Bernhard}, \citenamefont {Bessire}, \citenamefont {Feurer},\ and\ \citenamefont {Stefanov}}]{freq_ent}%
  \BibitemOpen
  \bibfield  {author} {\bibinfo {author} {\bibfnamefont {C.}~\bibnamefont {Bernhard}}, \bibinfo {author} {\bibfnamefont {B.}~\bibnamefont {Bessire}}, \bibinfo {author} {\bibfnamefont {T.}~\bibnamefont {Feurer}}, \ and\ \bibinfo {author} {\bibfnamefont {A.}~\bibnamefont {Stefanov}},\ }\bibfield  {title} {\enquote {\bibinfo {title} {Shaping frequency-entangled qudits},}\ }\href {\doibase 10.1103/PhysRevA.88.032322} {\bibfield  {journal} {\bibinfo  {journal} {Phys. Rev. A}\ }\textbf {\bibinfo {volume} {88}},\ \bibinfo {pages} {032322} (\bibinfo {year} {2013})}\BibitemShut {NoStop}%
\bibitem [{\citenamefont {Cuevas}\ \emph {et~al.}(2013)\citenamefont {Cuevas}, \citenamefont {Carvacho}, \citenamefont {Saavedra}, \citenamefont {Cari{\~n}e}, \citenamefont {Nogueira}, \citenamefont {Figueroa}, \citenamefont {Cabello}, \citenamefont {Mataloni}, \citenamefont {Lima},\ and\ \citenamefont {Xavier}}]{energy_time_ent}%
  \BibitemOpen
  \bibfield  {author} {\bibinfo {author} {\bibfnamefont {A.}~\bibnamefont {Cuevas}}, \bibinfo {author} {\bibfnamefont {G.}~\bibnamefont {Carvacho}}, \bibinfo {author} {\bibfnamefont {G.}~\bibnamefont {Saavedra}}, \bibinfo {author} {\bibfnamefont {J.}~\bibnamefont {Cari{\~n}e}}, \bibinfo {author} {\bibfnamefont {W.~a.~T.}\ \bibnamefont {Nogueira}}, \bibinfo {author} {\bibfnamefont {M.}~\bibnamefont {Figueroa}}, \bibinfo {author} {\bibfnamefont {A.}~\bibnamefont {Cabello}}, \bibinfo {author} {\bibfnamefont {P.}~\bibnamefont {Mataloni}}, \bibinfo {author} {\bibfnamefont {G.}~\bibnamefont {Lima}}, \ and\ \bibinfo {author} {\bibfnamefont {G.~B.}\ \bibnamefont {Xavier}},\ }\bibfield  {title} {\enquote {\bibinfo {title} {Long-distance distribution of genuine energy-time entanglement},}\ }\href {\doibase 10.1038/ncomms3871} {\bibfield  {journal} {\bibinfo  {journal} {Nature Communications}\ }\textbf {\bibinfo {volume} {4}},\ \bibinfo {pages} {2871} (\bibinfo {year} {2013})}\BibitemShut {NoStop}%
\bibitem [{\citenamefont {Ma}\ \emph {et~al.}(2009)\citenamefont {Ma}, \citenamefont {Slattery}, \citenamefont {Chang},\ and\ \citenamefont {Tang}}]{time_bin_ent}%
  \BibitemOpen
  \bibfield  {author} {\bibinfo {author} {\bibfnamefont {L.}~\bibnamefont {Ma}}, \bibinfo {author} {\bibfnamefont {O.}~\bibnamefont {Slattery}}, \bibinfo {author} {\bibfnamefont {T.}~\bibnamefont {Chang}}, \ and\ \bibinfo {author} {\bibfnamefont {X.}~\bibnamefont {Tang}},\ }\bibfield  {title} {\enquote {\bibinfo {title} {Non-degenerated sequential time-bin entanglement generation using periodically poled ktp waveguide},}\ }\href@noop {} {\bibfield  {journal} {\bibinfo  {journal} {Opt. Express}\ }\textbf {\bibinfo {volume} {17}},\ \bibinfo {pages} {15799--15807} (\bibinfo {year} {2009})}\BibitemShut {NoStop}%
\bibitem [{\citenamefont {Cao}\ \emph {et~al.}(2020)\citenamefont {Cao}, \citenamefont {Gao}, \citenamefont {Zhang}, \citenamefont {Wang}, \citenamefont {He}, \citenamefont {Liu}, \citenamefont {Zhou}, \citenamefont {Chen}, \citenamefont {Li}, \citenamefont {Yu}, \citenamefont {Romero}, \citenamefont {Huang}, \citenamefont {Li},\ and\ \citenamefont {Guo}}]{oam_entang}%
  \BibitemOpen
  \bibfield  {author} {\bibinfo {author} {\bibfnamefont {H.}~\bibnamefont {Cao}}, \bibinfo {author} {\bibfnamefont {S.-C.}\ \bibnamefont {Gao}}, \bibinfo {author} {\bibfnamefont {C.}~\bibnamefont {Zhang}}, \bibinfo {author} {\bibfnamefont {J.}~\bibnamefont {Wang}}, \bibinfo {author} {\bibfnamefont {D.-Y.}\ \bibnamefont {He}}, \bibinfo {author} {\bibfnamefont {B.-H.}\ \bibnamefont {Liu}}, \bibinfo {author} {\bibfnamefont {Z.-W.}\ \bibnamefont {Zhou}}, \bibinfo {author} {\bibfnamefont {Y.-J.}\ \bibnamefont {Chen}}, \bibinfo {author} {\bibfnamefont {Z.-H.}\ \bibnamefont {Li}}, \bibinfo {author} {\bibfnamefont {S.-Y.}\ \bibnamefont {Yu}}, \bibinfo {author} {\bibfnamefont {J.}~\bibnamefont {Romero}}, \bibinfo {author} {\bibfnamefont {Y.-F.}\ \bibnamefont {Huang}}, \bibinfo {author} {\bibfnamefont {C.-F.}\ \bibnamefont {Li}}, \ and\ \bibinfo {author} {\bibfnamefont {G.-C.}\ \bibnamefont {Guo}},\ }\bibfield  {title} {\enquote {\bibinfo {title} {Distribution of high-dimensional orbital angular momentum entanglement
  over a 1km few-mode fiber},}\ }\href {https://opg.optica.org/optica/abstract.cfm?URI=optica-7-3-232} {\bibfield  {journal} {\bibinfo  {journal} {Optica}\ }\textbf {\bibinfo {volume} {7}},\ \bibinfo {pages} {232--237} (\bibinfo {year} {2020})}\BibitemShut {NoStop}%
\bibitem [{\citenamefont {Zhao}\ \emph {et~al.}(2023)\citenamefont {Zhao}, \citenamefont {Yang}, \citenamefont {Zhu}, \citenamefont {Zhou}, \citenamefont {Zhong}, \citenamefont {Du},\ and\ \citenamefont {Sheng}}]{hyper_ent}%
  \BibitemOpen
  \bibfield  {author} {\bibinfo {author} {\bibfnamefont {P.}~\bibnamefont {Zhao}}, \bibinfo {author} {\bibfnamefont {M.-Y.}\ \bibnamefont {Yang}}, \bibinfo {author} {\bibfnamefont {S.}~\bibnamefont {Zhu}}, \bibinfo {author} {\bibfnamefont {L.}~\bibnamefont {Zhou}}, \bibinfo {author} {\bibfnamefont {W.}~\bibnamefont {Zhong}}, \bibinfo {author} {\bibfnamefont {M.-M.}\ \bibnamefont {Du}}, \ and\ \bibinfo {author} {\bibfnamefont {Y.-B.}\ \bibnamefont {Sheng}},\ }\bibfield  {title} {\enquote {\bibinfo {title} {Generation of hyperentangled state encoded in three degrees of freedom},}\ }\href {\doibase 10.1007/s11433-023-2164-7} {\bibfield  {journal} {\bibinfo  {journal} {Science China Physics, Mechanics \& Astronomy}\ }\textbf {\bibinfo {volume} {66}},\ \bibinfo {pages} {100311} (\bibinfo {year} {2023})}\BibitemShut {NoStop}%
\bibitem [{\citenamefont {Brambila}\ \emph {et~al.}(2023{\natexlab{a}})\citenamefont {Brambila}, \citenamefont {G\'{o}mez}, \citenamefont {Fazili}, \citenamefont {Gr\"{a}fe},\ and\ \citenamefont {Steinlechner}}]{pol_ent}%
  \BibitemOpen
  \bibfield  {author} {\bibinfo {author} {\bibfnamefont {E.}~\bibnamefont {Brambila}}, \bibinfo {author} {\bibfnamefont {R.}~\bibnamefont {G\'{o}mez}}, \bibinfo {author} {\bibfnamefont {R.}~\bibnamefont {Fazili}}, \bibinfo {author} {\bibfnamefont {M.}~\bibnamefont {Gr\"{a}fe}}, \ and\ \bibinfo {author} {\bibfnamefont {F.}~\bibnamefont {Steinlechner}},\ }\bibfield  {title} {\enquote {\bibinfo {title} {Ultrabright polarization-entangled photon pair source for frequency-multiplexed quantum communication in free-space},}\ }\href@noop {} {\bibfield  {journal} {\bibinfo  {journal} {Opt. Express}\ }\textbf {\bibinfo {volume} {31}},\ \bibinfo {pages} {16107--16117} (\bibinfo {year} {2023}{\natexlab{a}})}\BibitemShut {NoStop}%
\bibitem [{\citenamefont {Aspelmeyer}\ \emph {et~al.}(2003)\citenamefont {Aspelmeyer}, \citenamefont {Böhm}, \citenamefont {Gyatso}, \citenamefont {Jennewein}, \citenamefont {Kaltenbaek}, \citenamefont {Lindenthal}, \citenamefont {Molina-Terriza}, \citenamefont {Poppe}, \citenamefont {Resch}, \citenamefont {Taraba}, \citenamefont {Ursin}, \citenamefont {Walther},\ and\ \citenamefont {Zeilinger}}]{entang_aspelmeyer}%
  \BibitemOpen
  \bibfield  {author} {\bibinfo {author} {\bibfnamefont {M.}~\bibnamefont {Aspelmeyer}}, \bibinfo {author} {\bibfnamefont {H.~R.}\ \bibnamefont {Böhm}}, \bibinfo {author} {\bibfnamefont {T.}~\bibnamefont {Gyatso}}, \bibinfo {author} {\bibfnamefont {T.}~\bibnamefont {Jennewein}}, \bibinfo {author} {\bibfnamefont {R.}~\bibnamefont {Kaltenbaek}}, \bibinfo {author} {\bibfnamefont {M.}~\bibnamefont {Lindenthal}}, \bibinfo {author} {\bibfnamefont {G.}~\bibnamefont {Molina-Terriza}}, \bibinfo {author} {\bibfnamefont {A.}~\bibnamefont {Poppe}}, \bibinfo {author} {\bibfnamefont {K.}~\bibnamefont {Resch}}, \bibinfo {author} {\bibfnamefont {M.}~\bibnamefont {Taraba}}, \bibinfo {author} {\bibfnamefont {R.}~\bibnamefont {Ursin}}, \bibinfo {author} {\bibfnamefont {P.}~\bibnamefont {Walther}}, \ and\ \bibinfo {author} {\bibfnamefont {A.}~\bibnamefont {Zeilinger}},\ }\bibfield  {title} {\enquote {\bibinfo {title} {Long-distance free-space distribution of quantum entanglement},}\ }\href {\doibase 10.1126/science.1085593}
  {\bibfield  {journal} {\bibinfo  {journal} {Science}\ }\textbf {\bibinfo {volume} {301}},\ \bibinfo {pages} {621--623} (\bibinfo {year} {2003})}\BibitemShut {NoStop}%
\bibitem [{\citenamefont {Yin}\ \emph {et~al.}(2020)\citenamefont {Yin}, \citenamefont {Li}, \citenamefont {Liao}, \citenamefont {Yang}, \citenamefont {Cao}, \citenamefont {Zhang}, \citenamefont {Ren}, \citenamefont {Cai}, \citenamefont {Liu}, \citenamefont {Li}, \citenamefont {Shu}, \citenamefont {Huang}, \citenamefont {Deng}, \citenamefont {Li}, \citenamefont {Zhang}, \citenamefont {Liu}, \citenamefont {Chen}, \citenamefont {Lu}, \citenamefont {Wang}, \citenamefont {Xu}, \citenamefont {Wang}, \citenamefont {Peng}, \citenamefont {Ekert},\ and\ \citenamefont {Pan}}]{yinEntanglementbasedSecureQuantum2020a}%
  \BibitemOpen
  \bibfield  {author} {\bibinfo {author} {\bibfnamefont {J.}~\bibnamefont {Yin}}, \bibinfo {author} {\bibfnamefont {Y.-H.}\ \bibnamefont {Li}}, \bibinfo {author} {\bibfnamefont {S.-K.}\ \bibnamefont {Liao}}, \bibinfo {author} {\bibfnamefont {M.}~\bibnamefont {Yang}}, \bibinfo {author} {\bibfnamefont {Y.}~\bibnamefont {Cao}}, \bibinfo {author} {\bibfnamefont {L.}~\bibnamefont {Zhang}}, \bibinfo {author} {\bibfnamefont {J.-G.}\ \bibnamefont {Ren}}, \bibinfo {author} {\bibfnamefont {W.-Q.}\ \bibnamefont {Cai}}, \bibinfo {author} {\bibfnamefont {W.-Y.}\ \bibnamefont {Liu}}, \bibinfo {author} {\bibfnamefont {S.-L.}\ \bibnamefont {Li}}, \bibinfo {author} {\bibfnamefont {R.}~\bibnamefont {Shu}}, \bibinfo {author} {\bibfnamefont {Y.-M.}\ \bibnamefont {Huang}}, \bibinfo {author} {\bibfnamefont {L.}~\bibnamefont {Deng}}, \bibinfo {author} {\bibfnamefont {L.}~\bibnamefont {Li}}, \bibinfo {author} {\bibfnamefont {Q.}~\bibnamefont {Zhang}}, \bibinfo {author} {\bibfnamefont {N.-L.}\ \bibnamefont {Liu}}, \bibinfo {author}
  {\bibfnamefont {Y.-A.}\ \bibnamefont {Chen}}, \bibinfo {author} {\bibfnamefont {C.-Y.}\ \bibnamefont {Lu}}, \bibinfo {author} {\bibfnamefont {X.-B.}\ \bibnamefont {Wang}}, \bibinfo {author} {\bibfnamefont {F.}~\bibnamefont {Xu}}, \bibinfo {author} {\bibfnamefont {J.-Y.}\ \bibnamefont {Wang}}, \bibinfo {author} {\bibfnamefont {C.-Z.}\ \bibnamefont {Peng}}, \bibinfo {author} {\bibfnamefont {A.~K.}\ \bibnamefont {Ekert}}, \ and\ \bibinfo {author} {\bibfnamefont {J.-W.}\ \bibnamefont {Pan}},\ }\bibfield  {title} {\enquote {\bibinfo {title} {Entanglement-based secure quantum cryptography over 1,120 kilometres},}\ }\href {\doibase 10.1038/s41586-020-2401-y} {\bibfield  {journal} {\bibinfo  {journal} {Nature}\ }\textbf {\bibinfo {volume} {582}},\ \bibinfo {pages} {501--505} (\bibinfo {year} {2020})}\BibitemShut {NoStop}%
\bibitem [{\citenamefont {Anwar}\ \emph {et~al.}(2021)\citenamefont {Anwar}, \citenamefont {Perumangatt}, \citenamefont {Steinlechner}, \citenamefont {Jennewein},\ and\ \citenamefont {Ling}}]{SPDC_EPS_rev}%
  \BibitemOpen
  \bibfield  {author} {\bibinfo {author} {\bibfnamefont {A.}~\bibnamefont {Anwar}}, \bibinfo {author} {\bibfnamefont {C.}~\bibnamefont {Perumangatt}}, \bibinfo {author} {\bibfnamefont {F.}~\bibnamefont {Steinlechner}}, \bibinfo {author} {\bibfnamefont {T.}~\bibnamefont {Jennewein}}, \ and\ \bibinfo {author} {\bibfnamefont {A.}~\bibnamefont {Ling}},\ }\bibfield  {title} {\enquote {\bibinfo {title} {Entangled photon-pair sources based on three-wave mixing in bulk crystals},}\ }\href {\doibase 10.1063/5.0023103} {\bibfield  {journal} {\bibinfo  {journal} {Review of Scientific Instruments}\ }\textbf {\bibinfo {volume} {92}},\ \bibinfo {pages} {041101} (\bibinfo {year} {2021})}\BibitemShut {NoStop}%
\bibitem [{\citenamefont {Suhara}(2009)}]{waveguide_entang}%
  \BibitemOpen
  \bibfield  {author} {\bibinfo {author} {\bibfnamefont {T.}~\bibnamefont {Suhara}},\ }\bibfield  {title} {\enquote {\bibinfo {title} {Generation of quantum-entangled twin photons by waveguide nonlinear-optic devices},}\ }\href {\doibase https://doi.org/10.1002/lpor.200810054} {\bibfield  {journal} {\bibinfo  {journal} {Laser \& Photonics Reviews}\ }\textbf {\bibinfo {volume} {3}},\ \bibinfo {pages} {370--393} (\bibinfo {year} {2009})}\BibitemShut {NoStop}%
\bibitem [{\citenamefont {Schimpf}\ \emph {et~al.}(2021{\natexlab{a}})\citenamefont {Schimpf}, \citenamefont {Reindl}, \citenamefont {Basso~Basset}, \citenamefont {Jöns}, \citenamefont {Trotta},\ and\ \citenamefont {Rastelli}}]{qdots_3}%
  \BibitemOpen
  \bibfield  {author} {\bibinfo {author} {\bibfnamefont {C.}~\bibnamefont {Schimpf}}, \bibinfo {author} {\bibfnamefont {M.}~\bibnamefont {Reindl}}, \bibinfo {author} {\bibfnamefont {F.}~\bibnamefont {Basso~Basset}}, \bibinfo {author} {\bibfnamefont {K.~D.}\ \bibnamefont {Jöns}}, \bibinfo {author} {\bibfnamefont {R.}~\bibnamefont {Trotta}}, \ and\ \bibinfo {author} {\bibfnamefont {A.}~\bibnamefont {Rastelli}},\ }\bibfield  {title} {\enquote {\bibinfo {title} {Quantum dots as potential sources of strongly entangled photons: {Perspectives} and challenges for applications in quantum networks},}\ }\href {\doibase 10.1063/5.0038729} {\bibfield  {journal} {\bibinfo  {journal} {Applied Physics Letters}\ }\textbf {\bibinfo {volume} {118}},\ \bibinfo {pages} {100502} (\bibinfo {year} {2021}{\natexlab{a}})}\BibitemShut {NoStop}%
\bibitem [{\citenamefont {Politi}\ \emph {et~al.}(2008)\citenamefont {Politi}, \citenamefont {Cryan}, \citenamefont {Rarity}, \citenamefont {Yu},\ and\ \citenamefont {O'Brien}}]{silicon_PIC}%
  \BibitemOpen
  \bibfield  {author} {\bibinfo {author} {\bibfnamefont {A.}~\bibnamefont {Politi}}, \bibinfo {author} {\bibfnamefont {M.~J.}\ \bibnamefont {Cryan}}, \bibinfo {author} {\bibfnamefont {J.~G.}\ \bibnamefont {Rarity}}, \bibinfo {author} {\bibfnamefont {S.}~\bibnamefont {Yu}}, \ and\ \bibinfo {author} {\bibfnamefont {J.~L.}\ \bibnamefont {O'Brien}},\ }\bibfield  {title} {\enquote {\bibinfo {title} {Silica-on-silicon waveguide quantum circuits},}\ }\href {\doibase 10.1126/science.1155441} {\bibfield  {journal} {\bibinfo  {journal} {Science}\ }\textbf {\bibinfo {volume} {320}},\ \bibinfo {pages} {646--649} (\bibinfo {year} {2008})}\BibitemShut {NoStop}%
\bibitem [{\citenamefont {Kuo}, \citenamefont {Verma},\ and\ \citenamefont {Nam}(2020)}]{Kuo:20}%
  \BibitemOpen
  \bibfield  {author} {\bibinfo {author} {\bibfnamefont {P.~S.}\ \bibnamefont {Kuo}}, \bibinfo {author} {\bibfnamefont {V.~B.}\ \bibnamefont {Verma}}, \ and\ \bibinfo {author} {\bibfnamefont {S.~W.}\ \bibnamefont {Nam}},\ }\bibfield  {title} {\enquote {\bibinfo {title} {Demonstration of a polarization-entangled photon-pair source based on phase-modulated ppln},}\ }\href {\doibase 10.1364/OSAC.387449} {\bibfield  {journal} {\bibinfo  {journal} {OSA Continuum}\ }\textbf {\bibinfo {volume} {3}},\ \bibinfo {pages} {295--304} (\bibinfo {year} {2020})}\BibitemShut {NoStop}%
\bibitem [{\citenamefont {Mahmudlu}\ \emph {et~al.}(2023)\citenamefont {Mahmudlu}, \citenamefont {Johanning}, \citenamefont {{van Rees}}, \citenamefont {Khodadad~Kashi}, \citenamefont {Epping}, \citenamefont {Haldar}, \citenamefont {Boller},\ and\ \citenamefont {Kues}}]{mahmudluFullyOnchipPhotonic2023}%
  \BibitemOpen
  \bibfield  {author} {\bibinfo {author} {\bibfnamefont {H.}~\bibnamefont {Mahmudlu}}, \bibinfo {author} {\bibfnamefont {R.}~\bibnamefont {Johanning}}, \bibinfo {author} {\bibfnamefont {A.}~\bibnamefont {{van Rees}}}, \bibinfo {author} {\bibfnamefont {A.}~\bibnamefont {Khodadad~Kashi}}, \bibinfo {author} {\bibfnamefont {J.~P.}\ \bibnamefont {Epping}}, \bibinfo {author} {\bibfnamefont {R.}~\bibnamefont {Haldar}}, \bibinfo {author} {\bibfnamefont {K.-J.}\ \bibnamefont {Boller}}, \ and\ \bibinfo {author} {\bibfnamefont {M.}~\bibnamefont {Kues}},\ }\bibfield  {title} {\enquote {\bibinfo {title} {Fully on-chip photonic turnkey quantum source for entangled qubit/qudit state generation},}\ }\href {\doibase 10.1038/s41566-023-01193-1} {\bibfield  {journal} {\bibinfo  {journal} {Nature Photonics}\ }\textbf {\bibinfo {volume} {17}},\ \bibinfo {pages} {518--524} (\bibinfo {year} {2023})}\BibitemShut {NoStop}%
\bibitem [{\citenamefont {Achatz}\ \emph {et~al.}(2023)\citenamefont {Achatz}, \citenamefont {Bulla}, \citenamefont {Ecker}, \citenamefont {Ortega}, \citenamefont {Bartokos}, \citenamefont {{Alvarado-Zacarias}}, \citenamefont {{Amezcua-Correa}}, \citenamefont {Bohmann}, \citenamefont {Ursin},\ and\ \citenamefont {Huber}}]{achatz_2023}%
  \BibitemOpen
  \bibfield  {author} {\bibinfo {author} {\bibfnamefont {L.}~\bibnamefont {Achatz}}, \bibinfo {author} {\bibfnamefont {L.}~\bibnamefont {Bulla}}, \bibinfo {author} {\bibfnamefont {S.}~\bibnamefont {Ecker}}, \bibinfo {author} {\bibfnamefont {E.~A.}\ \bibnamefont {Ortega}}, \bibinfo {author} {\bibfnamefont {M.}~\bibnamefont {Bartokos}}, \bibinfo {author} {\bibfnamefont {J.~C.}\ \bibnamefont {{Alvarado-Zacarias}}}, \bibinfo {author} {\bibfnamefont {R.}~\bibnamefont {{Amezcua-Correa}}}, \bibinfo {author} {\bibfnamefont {M.}~\bibnamefont {Bohmann}}, \bibinfo {author} {\bibfnamefont {R.}~\bibnamefont {Ursin}}, \ and\ \bibinfo {author} {\bibfnamefont {M.}~\bibnamefont {Huber}},\ }\bibfield  {title} {\enquote {\bibinfo {title} {Simultaneous transmission of hyper-entanglement in three degrees of freedom through a multicore fiber},}\ }\href {\doibase 10.1038/s41534-023-00700-0} {\bibfield  {journal} {\bibinfo  {journal} {npj Quantum Information}\ }\textbf {\bibinfo {volume} {9}},\ \bibinfo {pages} {45} (\bibinfo {year}
  {2023})}\BibitemShut {NoStop}%
\bibitem [{\citenamefont {Wengerowsky}\ \emph {et~al.}(2018)\citenamefont {Wengerowsky}, \citenamefont {Joshi}, \citenamefont {Steinlechner}, \citenamefont {H{\"u}bel},\ and\ \citenamefont {Ursin}}]{WDM_2018}%
  \BibitemOpen
  \bibfield  {author} {\bibinfo {author} {\bibfnamefont {S.}~\bibnamefont {Wengerowsky}}, \bibinfo {author} {\bibfnamefont {S.~K.}\ \bibnamefont {Joshi}}, \bibinfo {author} {\bibfnamefont {F.}~\bibnamefont {Steinlechner}}, \bibinfo {author} {\bibfnamefont {H.}~\bibnamefont {H{\"u}bel}}, \ and\ \bibinfo {author} {\bibfnamefont {R.}~\bibnamefont {Ursin}},\ }\bibfield  {title} {\enquote {\bibinfo {title} {An entanglement-based wavelength-multiplexed quantum communication network},}\ }\href {\doibase 10.1038/s41586-018-0766-y} {\bibfield  {journal} {\bibinfo  {journal} {Nature}\ }\textbf {\bibinfo {volume} {564}},\ \bibinfo {pages} {225--228} (\bibinfo {year} {2018})}\BibitemShut {NoStop}%
\bibitem [{\citenamefont {Ortega}\ \emph {et~al.}(2021)\citenamefont {Ortega}, \citenamefont {Dovzhik}, \citenamefont {Fuenzalida}, \citenamefont {Wengerowsky}, \citenamefont {{Alvarado-Zacarias}}, \citenamefont {Shiozaki}, \citenamefont {{Amezcua-Correa}}, \citenamefont {Bohmann},\ and\ \citenamefont {Ursin}}]{ortega_2021}%
  \BibitemOpen
  \bibfield  {author} {\bibinfo {author} {\bibfnamefont {E.~A.}\ \bibnamefont {Ortega}}, \bibinfo {author} {\bibfnamefont {K.}~\bibnamefont {Dovzhik}}, \bibinfo {author} {\bibfnamefont {J.}~\bibnamefont {Fuenzalida}}, \bibinfo {author} {\bibfnamefont {S.}~\bibnamefont {Wengerowsky}}, \bibinfo {author} {\bibfnamefont {J.~C.}\ \bibnamefont {{Alvarado-Zacarias}}}, \bibinfo {author} {\bibfnamefont {R.~F.}\ \bibnamefont {Shiozaki}}, \bibinfo {author} {\bibfnamefont {R.}~\bibnamefont {{Amezcua-Correa}}}, \bibinfo {author} {\bibfnamefont {M.}~\bibnamefont {Bohmann}}, \ and\ \bibinfo {author} {\bibfnamefont {R.}~\bibnamefont {Ursin}},\ }\bibfield  {title} {\enquote {\bibinfo {title} {Experimental {{Space-Division Multiplexed Polarization-Entanglement Distribution}} through 12 {{Paths}} of a {{Multicore Fiber}}},}\ }\href {\doibase 10.1103/PRXQuantum.2.040356} {\bibfield  {journal} {\bibinfo  {journal} {PRX Quantum}\ }\textbf {\bibinfo {volume} {2}},\ \bibinfo {pages} {040356} (\bibinfo {year} {2021})}\BibitemShut
  {NoStop}%
\bibitem [{\citenamefont {Lohrmann}\ \emph {et~al.}(2020)\citenamefont {Lohrmann}, \citenamefont {Perumangatt}, \citenamefont {Villar},\ and\ \citenamefont {Ling}}]{linearDisp_EPS}%
  \BibitemOpen
  \bibfield  {author} {\bibinfo {author} {\bibfnamefont {A.}~\bibnamefont {Lohrmann}}, \bibinfo {author} {\bibfnamefont {C.}~\bibnamefont {Perumangatt}}, \bibinfo {author} {\bibfnamefont {A.}~\bibnamefont {Villar}}, \ and\ \bibinfo {author} {\bibfnamefont {A.}~\bibnamefont {Ling}},\ }\bibfield  {title} {\enquote {\bibinfo {title} {Broadband pumped polarization entangled photon-pair source in a linear beam displacement interferometer},}\ }\href {\doibase 10.1063/1.5124416} {\bibfield  {journal} {\bibinfo  {journal} {Applied Physics Letters}\ }\textbf {\bibinfo {volume} {116}},\ \bibinfo {pages} {021101} (\bibinfo {year} {2020})}\BibitemShut {NoStop}%
\bibitem [{\citenamefont {Horn}\ and\ \citenamefont {Jennewein}(2019)}]{dual_cryst_eps}%
  \BibitemOpen
  \bibfield  {author} {\bibinfo {author} {\bibfnamefont {R.}~\bibnamefont {Horn}}\ and\ \bibinfo {author} {\bibfnamefont {T.}~\bibnamefont {Jennewein}},\ }\bibfield  {title} {\enquote {\bibinfo {title} {Auto-balancing and robust interferometer designs for polarization entangled photon sources},}\ }\href {\doibase 10.1364/OE.27.017369} {\bibfield  {journal} {\bibinfo  {journal} {Opt. Express}\ }\textbf {\bibinfo {volume} {27}},\ \bibinfo {pages} {17369--17376} (\bibinfo {year} {2019})}\BibitemShut {NoStop}%
\bibitem [{\citenamefont {Perumangatt}, \citenamefont {Lohrmann},\ and\ \citenamefont {Ling}(2020)}]{linear_eps}%
  \BibitemOpen
  \bibfield  {author} {\bibinfo {author} {\bibfnamefont {C.}~\bibnamefont {Perumangatt}}, \bibinfo {author} {\bibfnamefont {A.}~\bibnamefont {Lohrmann}}, \ and\ \bibinfo {author} {\bibfnamefont {A.}~\bibnamefont {Ling}},\ }\bibfield  {title} {\enquote {\bibinfo {title} {Experimental conversion of position correlation into polarization entanglement},}\ }\href {\doibase 10.1103/PhysRevA.102.012404} {\bibfield  {journal} {\bibinfo  {journal} {Phys. Rev. A}\ }\textbf {\bibinfo {volume} {102}},\ \bibinfo {pages} {012404} (\bibinfo {year} {2020})}\BibitemShut {NoStop}%
\bibitem [{\citenamefont {Camphausen}\ \emph {et~al.}(2021)\citenamefont {Camphausen}, \citenamefont {Álvaro Cuevas}, \citenamefont {Duempelmann}, \citenamefont {Terborg}, \citenamefont {Wajs}, \citenamefont {Tisa}, \citenamefont {Ruggeri}, \citenamefont {Cusini}, \citenamefont {Steinlechner},\ and\ \citenamefont {Pruneri}}]{robin}%
  \BibitemOpen
  \bibfield  {author} {\bibinfo {author} {\bibfnamefont {R.}~\bibnamefont {Camphausen}}, \bibinfo {author} {\bibnamefont {Álvaro Cuevas}}, \bibinfo {author} {\bibfnamefont {L.}~\bibnamefont {Duempelmann}}, \bibinfo {author} {\bibfnamefont {R.~A.}\ \bibnamefont {Terborg}}, \bibinfo {author} {\bibfnamefont {E.}~\bibnamefont {Wajs}}, \bibinfo {author} {\bibfnamefont {S.}~\bibnamefont {Tisa}}, \bibinfo {author} {\bibfnamefont {A.}~\bibnamefont {Ruggeri}}, \bibinfo {author} {\bibfnamefont {I.}~\bibnamefont {Cusini}}, \bibinfo {author} {\bibfnamefont {F.}~\bibnamefont {Steinlechner}}, \ and\ \bibinfo {author} {\bibfnamefont {V.}~\bibnamefont {Pruneri}},\ }\bibfield  {title} {\enquote {\bibinfo {title} {A quantum-enhanced wide-field phase imager},}\ }\href {\doibase 10.1126/sciadv.abj2155} {\bibfield  {journal} {\bibinfo  {journal} {Science Advances}\ }\textbf {\bibinfo {volume} {7}},\ \bibinfo {pages} {eabj2155} (\bibinfo {year} {2021})}\BibitemShut {NoStop}%
\bibitem [{\citenamefont {Clauser}\ \emph {et~al.}(1969)\citenamefont {Clauser}, \citenamefont {Horne}, \citenamefont {Shimony},\ and\ \citenamefont {Holt}}]{CHSH}%
  \BibitemOpen
  \bibfield  {author} {\bibinfo {author} {\bibfnamefont {J.~F.}\ \bibnamefont {Clauser}}, \bibinfo {author} {\bibfnamefont {M.~A.}\ \bibnamefont {Horne}}, \bibinfo {author} {\bibfnamefont {A.}~\bibnamefont {Shimony}}, \ and\ \bibinfo {author} {\bibfnamefont {R.~A.}\ \bibnamefont {Holt}},\ }\bibfield  {title} {\enquote {\bibinfo {title} {Proposed experiment to test local hidden-variable theories},}\ }\href {\doibase 10.1103/PhysRevLett.23.880} {\bibfield  {journal} {\bibinfo  {journal} {Phys. Rev. Lett.}\ }\textbf {\bibinfo {volume} {23}},\ \bibinfo {pages} {880--884} (\bibinfo {year} {1969})}\BibitemShut {NoStop}%
\bibitem [{\citenamefont {Steinlechner}\ and\ \citenamefont {Pruneri}(2015)}]{steinlechnerSources2015}%
  \BibitemOpen
  \bibfield  {author} {\bibinfo {author} {\bibfnamefont {F.}~\bibnamefont {Steinlechner}}\ and\ \bibinfo {author} {\bibfnamefont {V.}~\bibnamefont {Pruneri}},\ }\emph {\bibinfo {title} {Sources of Photonic Entanglement for Applications in Space}},\ \href {\doibase 10.5821/dissertation-2117-96057} {Ph.D. thesis},\ \bibinfo  {school} {Universitat Polit{\`e}cnica de Catalunya} (\bibinfo {year} {2015})\BibitemShut {NoStop}%
\bibitem [{\citenamefont {Wengerowsky}\ and\ \citenamefont {Zeilinger}(2021)}]{wengerowskyExperiments}%
  \BibitemOpen
  \bibfield  {author} {\bibinfo {author} {\bibfnamefont {S.}~\bibnamefont {Wengerowsky}}\ and\ \bibinfo {author} {\bibfnamefont {A.}~\bibnamefont {Zeilinger}},\ }\emph {\bibinfo {title} {Two-Photon Polarization-Entanglement for Experiments and Applications in Quantum Communications}},\ \href@noop {} {Ph.D. thesis},\ \bibinfo  {school} {Universit{\"a}t Wien} (\bibinfo {year} {2021})\BibitemShut {NoStop}%
\bibitem [{\citenamefont {Kwiat}\ \emph {et~al.}(1999)\citenamefont {Kwiat}, \citenamefont {Waks}, \citenamefont {White}, \citenamefont {Appelbaum},\ and\ \citenamefont {Eberhard}}]{PhysRevA.60.R773}%
  \BibitemOpen
  \bibfield  {author} {\bibinfo {author} {\bibfnamefont {P.~G.}\ \bibnamefont {Kwiat}}, \bibinfo {author} {\bibfnamefont {E.}~\bibnamefont {Waks}}, \bibinfo {author} {\bibfnamefont {A.~G.}\ \bibnamefont {White}}, \bibinfo {author} {\bibfnamefont {I.}~\bibnamefont {Appelbaum}}, \ and\ \bibinfo {author} {\bibfnamefont {P.~H.}\ \bibnamefont {Eberhard}},\ }\bibfield  {title} {\enquote {\bibinfo {title} {Ultrabright source of polarization-entangled photons},}\ }\href {\doibase 10.1103/PhysRevA.60.R773} {\bibfield  {journal} {\bibinfo  {journal} {Phys. Rev. A}\ }\textbf {\bibinfo {volume} {60}},\ \bibinfo {pages} {R773--R776} (\bibinfo {year} {1999})}\BibitemShut {NoStop}%
\bibitem [{\citenamefont {Fiorentino}\ and\ \citenamefont {Beausoleil}(2008)}]{Comp_EPS_designs}%
  \BibitemOpen
  \bibfield  {author} {\bibinfo {author} {\bibfnamefont {M.}~\bibnamefont {Fiorentino}}\ and\ \bibinfo {author} {\bibfnamefont {R.~G.}\ \bibnamefont {Beausoleil}},\ }\bibfield  {title} {\enquote {\bibinfo {title} {Compact sources of polarization-entangled photons},}\ }\href {\doibase 10.1364/OE.16.020149} {\bibfield  {journal} {\bibinfo  {journal} {Opt. Express}\ }\textbf {\bibinfo {volume} {16}},\ \bibinfo {pages} {20149--20156} (\bibinfo {year} {2008})}\BibitemShut {NoStop}%
\bibitem [{\citenamefont {Aspect}(1976)}]{Aspect}%
  \BibitemOpen
  \bibfield  {author} {\bibinfo {author} {\bibfnamefont {A.}~\bibnamefont {Aspect}},\ }\bibfield  {title} {\enquote {\bibinfo {title} {Proposed experiment to test the nonseparability of quantum mechanics},}\ }\href {\doibase 10.1103/PhysRevD.14.1944} {\bibfield  {journal} {\bibinfo  {journal} {Phys. Rev. D}\ }\textbf {\bibinfo {volume} {14}},\ \bibinfo {pages} {1944--1951} (\bibinfo {year} {1976})}\BibitemShut {NoStop}%
\bibitem [{\citenamefont {Riedel~G{\aa}rding}\ \emph {et~al.}(2021)\citenamefont {Riedel~G{\aa}rding}, \citenamefont {Schwaller}, \citenamefont {Chan}, \citenamefont {Chang}, \citenamefont {Bosch}, \citenamefont {Gessler}, \citenamefont {Laborde}, \citenamefont {Hernandez}, \citenamefont {Si}, \citenamefont {Dupertuis},\ and\ \citenamefont {Macris}}]{riedelgardingBellDiagonalWerner2021}%
  \BibitemOpen
  \bibfield  {author} {\bibinfo {author} {\bibfnamefont {E.}~\bibnamefont {Riedel~G{\aa}rding}}, \bibinfo {author} {\bibfnamefont {N.}~\bibnamefont {Schwaller}}, \bibinfo {author} {\bibfnamefont {C.~L.}\ \bibnamefont {Chan}}, \bibinfo {author} {\bibfnamefont {S.~Y.}\ \bibnamefont {Chang}}, \bibinfo {author} {\bibfnamefont {S.}~\bibnamefont {Bosch}}, \bibinfo {author} {\bibfnamefont {F.}~\bibnamefont {Gessler}}, \bibinfo {author} {\bibfnamefont {W.~R.}\ \bibnamefont {Laborde}}, \bibinfo {author} {\bibfnamefont {J.~N.}\ \bibnamefont {Hernandez}}, \bibinfo {author} {\bibfnamefont {X.}~\bibnamefont {Si}}, \bibinfo {author} {\bibfnamefont {M.-A.}\ \bibnamefont {Dupertuis}}, \ and\ \bibinfo {author} {\bibfnamefont {N.}~\bibnamefont {Macris}},\ }\bibfield  {title} {\enquote {\bibinfo {title} {Bell {{Diagonal}} and {{Werner State Generation}}: {{Entanglement}}, {{Non-Locality}}, {{Steering}} and {{Discord}} on the {{IBM Quantum Computer}}},}\ }\href {\doibase 10.3390/e23070797} {\bibfield  {journal} {\bibinfo
  {journal} {Entropy}\ }\textbf {\bibinfo {volume} {23}},\ \bibinfo {pages} {797} (\bibinfo {year} {2021})}\BibitemShut {NoStop}%
\bibitem [{\citenamefont {Kim}, \citenamefont {Fiorentino},\ and\ \citenamefont {Wong}(2006)}]{PhysRevA.73.012316}%
  \BibitemOpen
  \bibfield  {author} {\bibinfo {author} {\bibfnamefont {T.}~\bibnamefont {Kim}}, \bibinfo {author} {\bibfnamefont {M.}~\bibnamefont {Fiorentino}}, \ and\ \bibinfo {author} {\bibfnamefont {F.~N.~C.}\ \bibnamefont {Wong}},\ }\bibfield  {title} {\enquote {\bibinfo {title} {Phase-stable source of polarization-entangled photons using a polarization sagnac interferometer},}\ }\href {\doibase 10.1103/PhysRevA.73.012316} {\bibfield  {journal} {\bibinfo  {journal} {Phys. Rev. A}\ }\textbf {\bibinfo {volume} {73}},\ \bibinfo {pages} {012316} (\bibinfo {year} {2006})}\BibitemShut {NoStop}%
\bibitem [{\citenamefont {Brambila}\ \emph {et~al.}(2023{\natexlab{b}})\citenamefont {Brambila}, \citenamefont {G{\'o}mez}, \citenamefont {Fazili}, \citenamefont {Gr{\"a}fe},\ and\ \citenamefont {Steinlechner}}]{brambila}%
  \BibitemOpen
  \bibfield  {author} {\bibinfo {author} {\bibfnamefont {E.}~\bibnamefont {Brambila}}, \bibinfo {author} {\bibfnamefont {R.}~\bibnamefont {G{\'o}mez}}, \bibinfo {author} {\bibfnamefont {R.}~\bibnamefont {Fazili}}, \bibinfo {author} {\bibfnamefont {M.}~\bibnamefont {Gr{\"a}fe}}, \ and\ \bibinfo {author} {\bibfnamefont {F.}~\bibnamefont {Steinlechner}},\ }\bibfield  {title} {\enquote {\bibinfo {title} {Ultrabright polarization-entangled photon pair source for frequency-multiplexed quantum communication in free-space},}\ }\href {\doibase 10.1364/OE.461802} {\bibfield  {journal} {\bibinfo  {journal} {Optics Express}\ }\textbf {\bibinfo {volume} {31}},\ \bibinfo {pages} {16107} (\bibinfo {year} {2023}{\natexlab{b}})}\BibitemShut {NoStop}%
\bibitem [{\citenamefont {Beckert}\ \emph {et~al.}(2019)\citenamefont {Beckert}, \citenamefont {de~Vries}, \citenamefont {Ursin}, \citenamefont {Steinlechner}, \citenamefont {Gr{\"a}fe},\ and\ \citenamefont {Basset}}]{beckert_2019}%
  \BibitemOpen
  \bibfield  {author} {\bibinfo {author} {\bibfnamefont {E.}~\bibnamefont {Beckert}}, \bibinfo {author} {\bibfnamefont {O.}~\bibnamefont {de~Vries}}, \bibinfo {author} {\bibfnamefont {R.}~\bibnamefont {Ursin}}, \bibinfo {author} {\bibfnamefont {F.-O.}\ \bibnamefont {Steinlechner}}, \bibinfo {author} {\bibfnamefont {M.}~\bibnamefont {Gr{\"a}fe}}, \ and\ \bibinfo {author} {\bibfnamefont {M.~G.}\ \bibnamefont {Basset}},\ }\bibfield  {title} {\enquote {\bibinfo {title} {A space-suitable, high brilliant entangled photon source for satellite based quantum key distribution},}\ \ }(\bibinfo  {publisher} {SPIE},\ \bibinfo {year} {2019})\ pp.\ \bibinfo {pages} {260--273}\BibitemShut {NoStop}%
\bibitem [{\citenamefont {Cao}\ \emph {et~al.}(2021)\citenamefont {Cao}, \citenamefont {Hisamitsu}, \citenamefont {Tokuda}, \citenamefont {Kurimura}, \citenamefont {Okamoto},\ and\ \citenamefont {Takeuchi}}]{chirped_SPDC}%
  \BibitemOpen
  \bibfield  {author} {\bibinfo {author} {\bibfnamefont {B.}~\bibnamefont {Cao}}, \bibinfo {author} {\bibfnamefont {M.}~\bibnamefont {Hisamitsu}}, \bibinfo {author} {\bibfnamefont {K.}~\bibnamefont {Tokuda}}, \bibinfo {author} {\bibfnamefont {S.}~\bibnamefont {Kurimura}}, \bibinfo {author} {\bibfnamefont {R.}~\bibnamefont {Okamoto}}, \ and\ \bibinfo {author} {\bibfnamefont {S.}~\bibnamefont {Takeuchi}},\ }\bibfield  {title} {\enquote {\bibinfo {title} {Efficient generation of ultra-broadband parametric fluorescence using chirped quasi-phase-matched waveguide devices},}\ }\href {\doibase 10.1364/OE.426575} {\bibfield  {journal} {\bibinfo  {journal} {Opt. Express}\ }\textbf {\bibinfo {volume} {29}},\ \bibinfo {pages} {21615--21628} (\bibinfo {year} {2021})}\BibitemShut {NoStop}%
\bibitem [{\citenamefont {Ecker}\ \emph {et~al.}(2021)\citenamefont {Ecker}, \citenamefont {Liu}, \citenamefont {Handsteiner}, \citenamefont {Fink}, \citenamefont {Rauch}, \citenamefont {Steinlechner}, \citenamefont {Scheidl}, \citenamefont {Zeilinger},\ and\ \citenamefont {Ursin}}]{eckerStrategiesAchievingHigh2021a}%
  \BibitemOpen
  \bibfield  {author} {\bibinfo {author} {\bibfnamefont {S.}~\bibnamefont {Ecker}}, \bibinfo {author} {\bibfnamefont {B.}~\bibnamefont {Liu}}, \bibinfo {author} {\bibfnamefont {J.}~\bibnamefont {Handsteiner}}, \bibinfo {author} {\bibfnamefont {M.}~\bibnamefont {Fink}}, \bibinfo {author} {\bibfnamefont {D.}~\bibnamefont {Rauch}}, \bibinfo {author} {\bibfnamefont {F.}~\bibnamefont {Steinlechner}}, \bibinfo {author} {\bibfnamefont {T.}~\bibnamefont {Scheidl}}, \bibinfo {author} {\bibfnamefont {A.}~\bibnamefont {Zeilinger}}, \ and\ \bibinfo {author} {\bibfnamefont {R.}~\bibnamefont {Ursin}},\ }\bibfield  {title} {\enquote {\bibinfo {title} {Strategies for achieving high key rates in satellite-based {{QKD}}},}\ }\href {\doibase 10.1038/s41534-020-00335-5} {\bibfield  {journal} {\bibinfo  {journal} {npj Quantum Information}\ }\textbf {\bibinfo {volume} {7}},\ \bibinfo {pages} {5} (\bibinfo {year} {2021})}\BibitemShut {NoStop}%
\bibitem [{\citenamefont {Steinlechner}\ \emph {et~al.}(2014)\citenamefont {Steinlechner}, \citenamefont {Gilaberte}, \citenamefont {Jofre}, \citenamefont {Scheidl}, \citenamefont {Torres}, \citenamefont {Pruneri},\ and\ \citenamefont {Ursin}}]{steinlechnerEfficientHeraldingPolarizationentangled2014b}%
  \BibitemOpen
  \bibfield  {author} {\bibinfo {author} {\bibfnamefont {F.}~\bibnamefont {Steinlechner}}, \bibinfo {author} {\bibfnamefont {M.}~\bibnamefont {Gilaberte}}, \bibinfo {author} {\bibfnamefont {M.}~\bibnamefont {Jofre}}, \bibinfo {author} {\bibfnamefont {T.}~\bibnamefont {Scheidl}}, \bibinfo {author} {\bibfnamefont {J.~P.}\ \bibnamefont {Torres}}, \bibinfo {author} {\bibfnamefont {V.}~\bibnamefont {Pruneri}}, \ and\ \bibinfo {author} {\bibfnamefont {R.}~\bibnamefont {Ursin}},\ }\bibfield  {title} {\enquote {\bibinfo {title} {Efficient heralding of polarization-entangled photons from type-0 and type-{{II}} spontaneous parametric downconversion in periodically poled {{KTiOPO}}\_4},}\ }\href {\doibase 10.1364/JOSAB.31.002068} {\bibfield  {journal} {\bibinfo  {journal} {Journal of the Optical Society of America B}\ }\textbf {\bibinfo {volume} {31}},\ \bibinfo {pages} {2068} (\bibinfo {year} {2014})}\BibitemShut {NoStop}%
\bibitem [{\citenamefont {Pomarico}\ \emph {et~al.}(2009)\citenamefont {Pomarico}, \citenamefont {Sanguinetti}, \citenamefont {Gisin}, \citenamefont {Thew}, \citenamefont {Zbinden}, \citenamefont {Schreiber}, \citenamefont {Thomas},\ and\ \citenamefont {Sohler}}]{waveg_OPO_ent}%
  \BibitemOpen
  \bibfield  {author} {\bibinfo {author} {\bibfnamefont {E.}~\bibnamefont {Pomarico}}, \bibinfo {author} {\bibfnamefont {B.}~\bibnamefont {Sanguinetti}}, \bibinfo {author} {\bibfnamefont {N.}~\bibnamefont {Gisin}}, \bibinfo {author} {\bibfnamefont {R.}~\bibnamefont {Thew}}, \bibinfo {author} {\bibfnamefont {H.}~\bibnamefont {Zbinden}}, \bibinfo {author} {\bibfnamefont {G.}~\bibnamefont {Schreiber}}, \bibinfo {author} {\bibfnamefont {A.}~\bibnamefont {Thomas}}, \ and\ \bibinfo {author} {\bibfnamefont {W.}~\bibnamefont {Sohler}},\ }\bibfield  {title} {\enquote {\bibinfo {title} {Waveguide-based opo source of entangled photon pairs},}\ }\href {\doibase 10.1088/1367-2630/11/11/113042} {\bibfield  {journal} {\bibinfo  {journal} {New Journal of Physics}\ }\textbf {\bibinfo {volume} {11}},\ \bibinfo {pages} {113042} (\bibinfo {year} {2009})}\BibitemShut {NoStop}%
\bibitem [{\citenamefont {Mower}\ and\ \citenamefont {Englund}(2011)}]{silicon_PIC2}%
  \BibitemOpen
  \bibfield  {author} {\bibinfo {author} {\bibfnamefont {J.}~\bibnamefont {Mower}}\ and\ \bibinfo {author} {\bibfnamefont {D.}~\bibnamefont {Englund}},\ }\bibfield  {title} {\enquote {\bibinfo {title} {Efficient generation of single and entangled photons on a silicon photonic integrated chip},}\ }\href {\doibase 10.1103/PhysRevA.84.052326} {\bibfield  {journal} {\bibinfo  {journal} {Phys. Rev. A}\ }\textbf {\bibinfo {volume} {84}},\ \bibinfo {pages} {052326} (\bibinfo {year} {2011})}\BibitemShut {NoStop}%
\bibitem [{\citenamefont {Vajner}\ \emph {et~al.}(2022)\citenamefont {Vajner}, \citenamefont {Rickert}, \citenamefont {Gao}, \citenamefont {Kaymazlar},\ and\ \citenamefont {Heindel}}]{qdots_rev}%
  \BibitemOpen
  \bibfield  {author} {\bibinfo {author} {\bibfnamefont {D.~A.}\ \bibnamefont {Vajner}}, \bibinfo {author} {\bibfnamefont {L.}~\bibnamefont {Rickert}}, \bibinfo {author} {\bibfnamefont {T.}~\bibnamefont {Gao}}, \bibinfo {author} {\bibfnamefont {K.}~\bibnamefont {Kaymazlar}}, \ and\ \bibinfo {author} {\bibfnamefont {T.}~\bibnamefont {Heindel}},\ }\bibfield  {title} {\enquote {\bibinfo {title} {Quantum {Communication} {Using} {Semiconductor} {Quantum} {Dots}},}\ }\href {\doibase 10.1002/qute.202100116} {\bibfield  {journal} {\bibinfo  {journal} {Advanced Quantum Technologies}\ }\textbf {\bibinfo {volume} {5}} (\bibinfo {year} {2022}),\ 10.1002/qute.202100116}\BibitemShut {NoStop}%
\bibitem [{\citenamefont {Hudson}\ \emph {et~al.}(2007)\citenamefont {Hudson}, \citenamefont {Stevenson}, \citenamefont {Bennett}, \citenamefont {Young}, \citenamefont {Nicoll}, \citenamefont {Atkinson}, \citenamefont {Cooper}, \citenamefont {Ritchie},\ and\ \citenamefont {Shields}}]{exciton_fidelity}%
  \BibitemOpen
  \bibfield  {author} {\bibinfo {author} {\bibfnamefont {A.~J.}\ \bibnamefont {Hudson}}, \bibinfo {author} {\bibfnamefont {R.~M.}\ \bibnamefont {Stevenson}}, \bibinfo {author} {\bibfnamefont {A.~J.}\ \bibnamefont {Bennett}}, \bibinfo {author} {\bibfnamefont {R.~J.}\ \bibnamefont {Young}}, \bibinfo {author} {\bibfnamefont {C.~A.}\ \bibnamefont {Nicoll}}, \bibinfo {author} {\bibfnamefont {P.}~\bibnamefont {Atkinson}}, \bibinfo {author} {\bibfnamefont {K.}~\bibnamefont {Cooper}}, \bibinfo {author} {\bibfnamefont {D.~A.}\ \bibnamefont {Ritchie}}, \ and\ \bibinfo {author} {\bibfnamefont {A.~J.}\ \bibnamefont {Shields}},\ }\bibfield  {title} {\enquote {\bibinfo {title} {Coherence of an entangled exciton-photon state},}\ }\href {\doibase 10.1103/PhysRevLett.99.266802} {\bibfield  {journal} {\bibinfo  {journal} {Phys. Rev. Lett.}\ }\textbf {\bibinfo {volume} {99}},\ \bibinfo {pages} {266802} (\bibinfo {year} {2007})}\BibitemShut {NoStop}%
\bibitem [{\citenamefont {Huber}\ \emph {et~al.}(2018)\citenamefont {Huber}, \citenamefont {Reindl}, \citenamefont {Aberl}, \citenamefont {Rastelli},\ and\ \citenamefont {Trotta}}]{qdots}%
  \BibitemOpen
  \bibfield  {author} {\bibinfo {author} {\bibfnamefont {D.}~\bibnamefont {Huber}}, \bibinfo {author} {\bibfnamefont {M.}~\bibnamefont {Reindl}}, \bibinfo {author} {\bibfnamefont {J.}~\bibnamefont {Aberl}}, \bibinfo {author} {\bibfnamefont {A.}~\bibnamefont {Rastelli}}, \ and\ \bibinfo {author} {\bibfnamefont {R.}~\bibnamefont {Trotta}},\ }\bibfield  {title} {\enquote {\bibinfo {title} {Semiconductor quantum dots as an ideal source of polarization-entangled photon pairs on-demand: a review},}\ }\href {\doibase 10.1088/2040-8986/aac4c4} {\bibfield  {journal} {\bibinfo  {journal} {Journal of Optics}\ }\textbf {\bibinfo {volume} {20}},\ \bibinfo {pages} {073002} (\bibinfo {year} {2018})}\BibitemShut {NoStop}%
\bibitem [{\citenamefont {Schimpf}\ \emph {et~al.}(2021{\natexlab{b}})\citenamefont {Schimpf}, \citenamefont {Reindl}, \citenamefont {Basso~Basset}, \citenamefont {Jöns}, \citenamefont {Trotta},\ and\ \citenamefont {Rastelli}}]{qdots_2}%
  \BibitemOpen
  \bibfield  {author} {\bibinfo {author} {\bibfnamefont {C.}~\bibnamefont {Schimpf}}, \bibinfo {author} {\bibfnamefont {M.}~\bibnamefont {Reindl}}, \bibinfo {author} {\bibfnamefont {F.}~\bibnamefont {Basso~Basset}}, \bibinfo {author} {\bibfnamefont {K.~D.}\ \bibnamefont {Jöns}}, \bibinfo {author} {\bibfnamefont {R.}~\bibnamefont {Trotta}}, \ and\ \bibinfo {author} {\bibfnamefont {A.}~\bibnamefont {Rastelli}},\ }\bibfield  {title} {\enquote {\bibinfo {title} {{Quantum dots as potential sources of strongly entangled photons: Perspectives and challenges for applications in quantum networks}},}\ }\href {\doibase 10.1063/5.0038729} {\bibfield  {journal} {\bibinfo  {journal} {Applied Physics Letters}\ }\textbf {\bibinfo {volume} {118}},\ \bibinfo {pages} {100502} (\bibinfo {year} {2021}{\natexlab{b}})}\BibitemShut {NoStop}%
\bibitem [{\citenamefont {Bell}(1964)}]{Bell}%
  \BibitemOpen
  \bibfield  {author} {\bibinfo {author} {\bibfnamefont {J.~S.}\ \bibnamefont {Bell}},\ }\bibfield  {title} {\enquote {\bibinfo {title} {On the einstein podolsky rosen paradox},}\ }\href {\doibase 10.1103/PhysicsPhysiqueFizika.1.195} {\bibfield  {journal} {\bibinfo  {journal} {Physics Physique Fizika}\ }\textbf {\bibinfo {volume} {1}},\ \bibinfo {pages} {195--200} (\bibinfo {year} {1964})}\BibitemShut {NoStop}%
\end{thebibliography}%

\end{document}

\preprint{AIP/123-QED}

\title[]{Supplement 1}

\maketitle

\subsection{Mechanical Sensitivity}
\label{sec:mechanical_sensitivity}

In a standard BD (lateral displacement on one polarization only), the shear equals $S_\text{BD}=\delta_\mathrm{BD}$ , while in a SP (perpendicular lateral displacement between both polarizations) the shear equals $S_\text{SP}=\sqrt{2}\delta_\mathrm{SP}$. For $S_\text{SP}=S_\text{BD}$, we get $\delta_\mathrm{SP}=\delta_\mathrm{BD}/\sqrt{2}$. Since the shear is linear with the components' thickness and the SP is essentially composed of two BDs, one can write $\frac{Z_\mathrm{SP}}{Z_\mathrm{BD}}=2(\frac{\delta_\mathrm{SP}}{\delta_\mathrm{BD}})=\sqrt{2}$.

In order to measure the robustness of our design versus state-of-the-art EPSs, we prepared a test to compare the phase response $\varphi$ analyzed in appendix A versus the tilting angle $\theta$ of the beam displacement component. In particular, we compared a Savart plate (SP) with a standard beam displacer (BD), both made on calcite and for a lateral shear of $S=\SI{1.2}{mm}$ on a wavelength $\SI{775}{nm}$. The SP and the BD were placed one at the time on a motorized tilting stage with a rotation angle that maximized $\varphi$ respect to input polarization, fixed by a polarizer. The light source consisted in a long coherence length laser (DLL-Toptica), which was injected into the setup from a single-mode fiber to a $ f = \SI{4}{mm} $ reflective collimator. The above ensured large overlap between the split beams either spatially and temporarily. Finally, a polarization CMOS camera was used for data acquisition.

\begin{figure}[H]
\begin{center}
\includegraphics[scale=0.24]{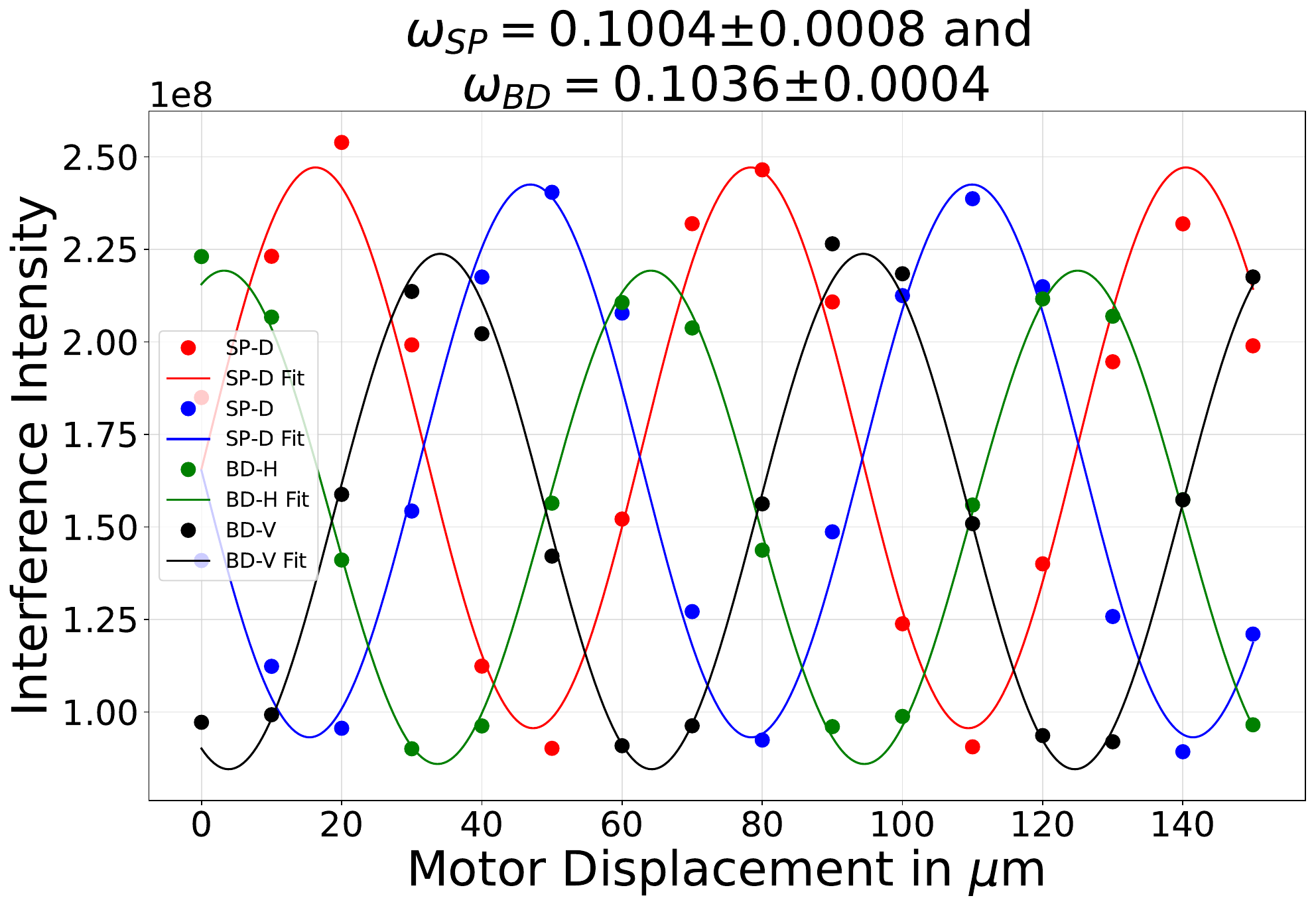}
\caption{\label{fig:phase_comparison}\textbf{Comparison of phase sensitivity vs tilting}. Light intensity over the camera interference quadrants, while scanning the phase $\varphi$ as a function of the motor displacement ($x$). In average, the SP (BD) manifests an interference frequency of $\omega_\mathrm{SP}=0.1004\pm0.0008$ ($\omega_\mathrm{SP}=0.1036\pm0.0004$), which for this data indicates a difference beyond the fitting errors. The tilting angle can be calculated as $\theta=\arctan(x/x_{p})$, where $x_{p}\approx \SI{8}{cm}$ is the pivot distance of the tilting stage.}
\end{center}
\end{figure}

In Fig. \ref{fig:phase_comparison} we show the detected cross section of the interference profile between the two beams, where the interference fringes are fitted via a standard cosine $\varphi=a\cos(\omega x+c)+d$ and the frequencies $\omega_{\mathrm{SP}}=0.1004\pm0.0008$ and $\omega_{\mathrm{BD}}=0.1036\pm0.0004 \,\text{rad/s}$ are obtained. We can therefore affirm that with a SP that is $\sqrt{2}$ thicker than a standard BD, the phase-shift between the two beams is less sensitive to the tilting angle. 

